\documentclass[twocolumn,secnumarabic,tightenlines,amssymb, amsmath,nobibnotes, aps]{revtex4}
\pdfoutput=1
\usepackage{docs}%
\usepackage{bm}%
\usepackage{graphicx}
\usepackage{amssymb}
\usepackage{amsthm,setspace}
\expandafter\ifx\csname package@font\endcsname\relax\else
 \expandafter\expandafter
 \expandafter\usepackage
 \expandafter\expandafter
 \expandafter{\csname package@font\endcsname}%
\fi

\begin{document}

\title{Nonlinear dynamics experiments in plasmas }%

\author{Md. Nurujjaman}%
\email{jaman_nonlinear@yahoo.co.in}
\affiliation{Saha Institute of Nuclear Physics, Kolkata, India.\\
     A thesis submitted to \\the Board of Studies in Physical Sciences \\In partial fulfillment of the requirements\\
       For the Degree of\\ DOCTOR OF PHILOSOPHY \\of \\HOMI BHABHA NATIONAL INSTITUTE\\
                       INDIA}
\date{2009}%
\begin{abstract}

This thesis presents some results of the nonlinear dynamics experiments in an argon dc glow discharge plasma. The thesis has been divided into six Sections.

Section 1 contains an introduction to nonlinear dynamics experiments  in plasmas, and the time series analysis. The study of nonlinear dynamics or chaos theory has emerged in the last three decades or so as an important interdisciplinary area of research encompassing a wide range of fields like: fluids, plasmas, biomedical sciences, finance, turbulence, astronomy, material sciences, etc. The advent of fast computing systems have been of great help in solving nonlinear equations which were intractable analytically. Low dimensional deterministic chaos has become a part and parcel of almost every field of engineering like, mechanical, electrical, civil, chemical, etc., and this has lead to the development  of new tools and data analysis techniques, like nonlinear time series analysis, wavelet transform, etc.

  In an electron beam plasma system in 1984, Boswell  showed that the natural oscillations of an electron beam propagating parallel to a magnetic field, went through a period doubling bifurcation to a chaotic state with increase in the beam current.  In 1989, the intermittency route to a chaos was observed in the low frequency self oscillations in the undriven DC discharge system. Homoclinic chaos was reported in  the same electrical discharge plasma system in which the deterministic chaos had been observed by T. Braun, \emph{et al.} Period adding route to chaos and period subtracting had been observed in on ion beam plasma in a double plasma device. Other nonlinear phenomena like mode locking, period pulling, etc., had been observed by Klinger et al. In many other experiments almost a similar phenomena had been observed where different types of gases,  geometric configurations and  parametric regimes were explored.

  In all the above experiments, the common feature was that  the plasma dynamics goes towards a chaotic state as the discharge current or  voltage (DV) was raised. On the other hand, the inverse route, i.e., chaos to ordered state transition has not been observed experimentally in plasmas. The possibility of such a route in a plasma sheath had been shown by Tomejiro Yamagishi and Makoto Tsukayama [Tomejiro Yamagishi and Makoto Tsukayama, J. Phys. Soc. Jpn. 69, 2883 (2000)].  numerically, in the presence of dissipation. Generally, with dissipation or decrease in the input energy, the system goes to an ordered state. In our experiments the chaotic behavior of the floating potential fluctuations, goes to a regular state with increasing DV.

 In Section 2, we have described the experimental setup, diagnostics, and data analysis procedures. The experiments were preformed in an argon dc glow discharge plasma. Plasma was produced inside a cylindrical hollow cathode dc glow discharge system of radius 45 mm and cylindrical anode of radius $\approx1$ mm placed inside the cathode concentrically. The entire electrode system was housed inside a vacuum chamber whose pressure (p) could be varied between 0.001 to 1 bar. A dc discharge voltage applied between the two electrodes  produced the  plasma, and a Langmuir probe located between the anode and the cathode was used to monitor the floating potential fluctuations, and a black and white CCD camera was used to measure the size of the anode glow. The noise  (HP33120A) and a signal generators (Fluke PM5138A) were coupled with the discharge power supply for the noise induced experiments. The measured plasma density and the electron temperature were about $10^7cm^{-3}$ and $3-4$ eV respectively, and the corresponding electron plasma frequency ($f_{pe}$) and the ion plasma frequency ($f_{pi}$) were about 28 MHz and 105 kHz respectively. The analysis of the floating potential fluctuations has been carried out using the linear and nonlinear time series analysis. In the  linear analysis, power spectrum, autocorrelation, probability density function, etc.,  and in nonlinear analysis Lyapunov exponent, correlation dimension, Hurst exponent, etc., have been estimated. Some statistical tools like `normal variance' and `absolute mean difference' have also been used to characterize coherence and stochastic resonances. Most of the data analysis were carried out using the software MATLAB.

In Section 3, we have presented anode glow related observation of chaos to order transition and homoclinic bifurcation in the glow discharge plasma. Depending upon the gas pressure (p), a discharge was struck at different discharge voltages (DV), which exhibits a Paschen curve like behavior. The region to the left of the Paschen minimum is narrower than the region to the right and the behavior of the plasma floating potential fluctuations was different on these two sides of the Paschen curve. To right of  the Paschen minimum, an anode glow or anode spot was observed to form on the anode, whose size decreased  with the increase in the DV, before finally disappearing at a certain DV. The annular radii of the glow  estimated from the experimental observations  was in the range of $1.3-0.32$ mm.   This  is consistent with the theoretical estimations of thickness ($\delta$) using the relation $\delta\approx3.7\times10^{-6}\frac{kT}{\sigma_{i}P}$, where k, T, $\sigma_{i}$, and P are the Boltzmann constant, room temperature in Kelvin scale, ionization cross-section in $cm^{-2}$ and pressure in mbar respectively. The estimated thickness of the anode glow using the above relation for $P\approx0.95$ mbar, $T\approx300$ K, is $ \approx0.81$ mm is  within the range of the thickness as obtained  from the experiment ($1.3-0.32$ mm).

The floating potential fluctuations have been analyzed for three typical pressures in the region greater than the Paschen minimum. At about 0.89 mbar ($pd\approx 20.02$ mbar-mm), the discharge was initiated at $\approx 288$ V and simultaneously irregular relaxation type of oscillations were observed  but turned into regular oscillations with increase in the DVs. Around 509 V,  these oscillations disappeared and this point (DV) is termed as the bifurcation point ($V_H$). A similar behaviour was observed at 0.95 mbar ($pd=21.37$ mbar mm)  and at 1 mbar ($pd= 22.5$ mbar mm ). Initially, the power spectrum of the oscillations are of broad band nature indicating a chaotic behavior of the system.  An estimate of the frequency of these instabilities from the ion transit time in the plasma $\tau(d)=\frac{d}{V_{th,i}}=d/\sqrt\frac{k_bT_i}{m}$, where d is the electrode distance, is $\approx19$ kHz which agrees quite well with the frequency of the relaxation oscillations of the floating potential.

 The relaxation oscillations have been attributed to the formation of highly nonlinear structures like double layers, and hence, we tried to estimate the correlation dimension ($D_{corr}$) and the +ve Lyapunov exponent ($\lambda_L$) of all the signals.  The estimated $D_{corr}$ for the three pressures (0.89, 0.95 and 1.0 mbar) were greater than 3.8 to begin with and decreased with increase in DV. Since $D_{corr}$ is a measure of the complexity of the system, initially, the system was in a complex state as $D_{corr}$ for all the three pressures were high and decreased with increase in DV.The presence of a +ve Lyapunov exponent ($\lambda_L$) is the most reliable signature of the chaotic dynamics.  Positive $\lambda_L$ was obtained  for DV  283, 284, and 290 V at 0.95 mbar and for 293, 296, 300 and 305 V at 1 mbar respectively.  At higher pressures we find that in general $\lambda_L$ is +ve and $D_{corr}  \geq 3$, suggesting a low dimensional chaos. Wavelet analysis also showed a presence of chaos at the initial stage of the discharge.

Generally, the floating potential fluctuations were complex in nature at the initial stage of discharge and became regular with increase in DV and converted to relaxation oscillations before totally vanishing  at the bifurcation point ($V_H$).

The time period (T) of these relaxation oscillations increases  with increasing DV, before eventually vanishing  beyond a critical DV ($V_H$), and for larger values, the autonomous dynamics exhibits a steady state fixed point behavior. It is observed that the autonomous dynamics undergoes an exponential slowing down. It is observed that $\ln|V-V_H|$ vs Time is a straight line, indicating that the system goes through a homoclinic bifurcation.  The floating potential fluctuations exhibit relaxation oscillations on the one side of the $V_H$ (bifurcation point) and a stable fixed point on the other side,  called the excitable state and is useful to study noise invoked dynamics$-$ stochastic and coherence resonances.

In Section 4, the noise invoked dynamics has been investigated. When the DV is modulated with a stochastic signal, the autonomous dynamics will be recovered and this phenomena is termed as coherence resonance.  For stochastic resonance, the DV was perturbed by a combination of a subthreshold periodic pulse train  and a Gaussian white noise. The time series of the system response  in the presence of an identical subthreshold signal for three different amplitudes of imposed noise. It is observed that there is little correspondence between the subthreshold signal and the system response for a low noise amplitude. However, there is an excellent correspondence at an intermediate noise
amplitude. Finally,  at higher
amplitudes of noise, the subthreshold signal is lost amidst
stochastic fluctuations of the system
response. Absolute
mean difference (AMD) has been used to quantify the
information transfer between the subthreshold
signal and the system response and the unimodal structure is the signature of the SR phenomena.

 In Sections 5, we have presented the observation self organized criticality behavior in plasma.  When the system was operated in the pressure region less than the Paschen minimum, for a small range of p ($0.9-1.5\times10^{-2}$ Torr or $1.2-2\times10^{-2}$ mbar), it was observed that the behavior of the floating potential fluctuations was consistent with self organized criticality (SOC). In order to establish the SOC behavior, we had checked the power law behavior of the power spectrum, and the presence of the long-range correlation by estimating Hurst exponent (H) (using R/S technique) and the exponent ($\alpha$) of autocorrelation function (ACF) decay, and the nongaussian probability distribution function. The results of the Hurst exponent ($0.96\pm0.01$) being greater than 0.5, the ACF exponent, $\alpha\sim0.30$, nongaussian PDF, and power spectral index $\beta \approx 1.7\pm0.1$ in the pressure range $9\times10^{-3}-1.6\times10^{-2}$ Torr, are consistent with the systems exhibiting SOC like behavior.

Finally, we have presented a summary and future plan of  work in Section 6. Though we have investigated some very important and interesting  nonlinear processes  during the the course of this thesis work, there are, many problems which we plan to take up as a part of our future activities, are as follows.
\begin{itemize}
  \item Effect of the noise on the autonomous dynamics.
  \item Effect of suprathreshold signal and noise to the plasma.
  \item Investigation of the existence of canard orbits in plasma.
  \item Non-chaotic attractors may be investigated by applying two non-commensurate periodic signals.
      \item Chaos control and synchronization.
      \item Modeling of the experimental results.
\end{itemize}

Dusty plasma which can be produced very easily in our experimental system, is another area where, a lot of nonlinear phenomena  can be explored.

\end{abstract}
\maketitle
\tableofcontents

\section{Overview of nonlinear dynamics with relevance to plasma physics }
\label{section:introduction}
\subsection{Introduction}
Nonlinear dynamics and chaos theory  started with the intention of investigating the qualitative behavior of nonlinear problems which were  difficult to solve analytically. In the first half of the 1900s, scientists were interested in nonlinear oscillators like the van der Pol oscillator, and with the discovery of the high speed computers many different nonlinear problems were solved~\cite{book:strogatz}. In 1963, Lorenz~\cite{book:lakshmanan} introduced, the fluid convection model  to study the dynamics of the atmospheric weather, which exhibited chaotic behavior in the numerical simulations  and this model later came to be known as the famous Lorenz model. It shows very complicated dynamics like chaos, inverse period doubling bifurcation, etc., depending on the model parameters.  Chaos theory flourished when Reuelle and Takens~\cite{commathphys:ruelle} proposed a new theory for the onset of turbulence in the fluids and Gollub and Swinney verified it experimentally in a concentric rotating cylinder using water, and it was observed that with the increase in the rotation rate the radial velocity fluctuations went to a chaotic state through a period doubling bifurcation, which were analyzed using Fourier techniques~\cite{prl:gollub}. In the late 70s, Feigenbaum discovered the universal constant called the Feigenbaum constant to characterize the universal features of the period doubling bifurcation~\cite{jstatphys:feigenbaum}. Later, bifurcation diagrams, Lyapunov exponent, correlation dimension, etc., derived on the basis of chaos theory, have  been used to characterize chaos and its different routes~\cite{physrevlett:grassberger,PhysicaD:wolf,physicaD:rosenstein}. Since the development of nonlinear dynamics, most applications have been in the field of fluid dynamics, particularly fluid turbulence and since fluid and plasma turbulence are closely related, the concepts of nonlinear dynamics had been successfully utilized  in plasmas.

 In the late sixties and later, the idea of van der Pol oscillator was applied to explain the growth and saturation of the plasma instabilities  which could not be done using conventional linear theories~\cite{jphysDAppl:keen,physFluids:stix,physFluids:boswell,prl:deneef,physrev:Hsuan,physfluids:hendel,prl:keen,physfluids:boswell,pop:pecseli,pla:klinger,pre:klinger,pla:Gyergyek,pla:Buragohain,pla:Tavzes,pop:Klostermann,pop:Rohde,pop:Koepke,pre:Wendt,pscr:Klinger}.  Abrams \emph{et al.}, first observed the nonlinear phenomena like period pulling, frequency entrainment, etc., in periodically forced self oscillatory plasmas~\cite{prl:abrams}. B.E. Keen, \emph{et al.}, showed using a two fluid model that the ion sound instabilities behaved in a manner similar to a van der Pol oscillator when subjected to a driving force~\cite{jphysDAppl:keen}.  By the mid eighties researchers began looking for experimental evidences of deterministic chaos in plasmas. Experimentally, the period doubling bifurcation and chaotic behavior were observed in an electron beam plasma system in 1984, wherein, Boswell  showed that the natural oscillations on an electron beam propagating parallel to a magnetic field, went through a period doubling bifurcation to a chaotic state with increase in the beam current~\cite{ppcf:boswell}.  The period doubling route to chaos was also reported in the driven pulsed filament discharge plasma  and an electrical discharge tube~\cite{prl:Cheung,prl:braun}. In 1989, the intermittency route to a chaos was observed in the low frequency self oscillations in the undriven DC discharge system~\cite{chinesepl:yong,prl:Qin}. Homoclinic chaos~\cite{prl:braun1} was reported in  the same electrical discharge plasma system in which the deterministic chaos had been observed by T. Braun, \emph{et al.},  Quasiperiodic route to chaos was observed in the driven and undriven discharge plasmas~\cite{prl:Qin,physletta:Fan,prl:Weixing} in  magnetic multipole devices~\cite{physletta:Fan,prl:Weixing}. Period adding route to chaos and period subtracting had been observed in ion beam plasma in double plasma device~\cite{ijbc:sharma}. Other nonlinear phenomena like mode locking, period pulling, frequency entrainments etc., had been observed by Klinger et al~\cite{pop:Klinger,pla:klinger,pre:klinger,prl:abrams,georl:Koepke,book:Koepke,book:Koepke1,georl:Lashinsky}. In arc plasma  chaos have been observed by S. Ghorui, \emph{et al.},~\cite{pre:ghorui,pre:ghorui1,pre:ghorui2}. In many other experiments almost similar phenomena had been observed where different types of gases,  geometric configurations and  parametric regimes were explored~\cite{prl:Ding,IEEE:ref4,pop:Hassouba,EurPhysJD:Atipo,pop:lee1}. In all the above experiments the common feature was that  the plasma dynamics goes towards a chaotic state as the discharge current or discharge voltage (DV) was raised.

On the other hand, inverse route, i.e., chaos to ordered state transition has not been observed experimentally in plasmas. Possibility of such a route in plasma sheath had been shown by Tomejiro Yamagishi and Makoto Tsukayama~\cite{jpsj:makato1}  numerically, where dissipation was important to lead to such situations. Generally, with dissipation or decrease in the input energy, the system goes to an ordered state. In our experiments the chaotic behavior of the floating potential fluctuations generated during the creation of an anode glow, goes to a regular state with increasing the discharge voltage. This happens because with increase in DV, the glow vanishes and hence the system becomes stable and this has been discussed in detail in Section~\ref{section:chaos}.

With the advent of nonlinear dynamics, many other interesting  phenomena like controlling chaos~\cite{prl:ott}, noise induced resonances~\cite{JPhysA:benzi1}, synchronization of chaotic system~\cite{prl:pecora}, quantum chaos, etc., have been observed. Some of these phenomena are also observed in plasma, specially coherence resonance near Hopf bifurcation was studied  by Lin I and Jeng-Mei Liu~\cite{prl:Lin I1} and A. Dinklage, \emph{et al.}~\cite{pop:dinklage1}. Our experiments have been very conducive for investigation both stochastic and coherence resonance phenomena and have been discussed in Section~\ref{section:resonance}.   We are presenting results of both types of resonances in the glow discharge plasmas near homoclinic bifurcation in Section~\ref{section:resonance}.

\subsection{Time series analysis}
\label{subsection:time series}
The instabilities that we have studied in this thesis were recorded in the form of time dependent  floating potential fluctuations. To understand the characteristics of the instabilities, one has to analyze the fluctuations using the time series analysis which has been given here.

   A \emph{\textbf{time series}} is a sequence of data points of an observed variable at equally spaced time intervals and \emph{\textbf{time series analysis }} comprises of methods that attempt to understand such time series, often either to understand the underlying context of the data points (where did they come from? what generated them?), or to make forecasts (predictions). Applications cover virtually all areas of statistics, but some of the most important ones include economic and financial time series, analyzing and modeling and forecasting of physical time series (hydrology, earth sciences, astronomy, oceanography and biology, astrophysics, plasma, etc.), simulation of time series models and examination of their properties, filtering and smoothing of any type of time series, statistical estimation of parameters of the time series models, either linear or nonlinear, with several types of optimization methods, etc.

 The objectives of a time series analysis are diverse, depending on the background of applications. Statisticians whose fundamental task is to unveil the probability law that governs the observed time series, usually view them as a realization from a stochastic process, With such a probability law, we can understand the underlying dynamics, forecast future events, and control them via suitable intervention. Time series analysis can be categorized in two classes: linear  and nonlinear time series analysis.  Methods for linear time series analysis are often divided into two classes: frequency-domain and time-domain methods. The former are spectral and wavelet analysis while the latter consists of estimating the auto-correlation and cross-correlation in the time series.

Linear analysis tools treat all the irregular or chaotic behavior as a stochastic processes though these instabilities may be generated by deterministic dynamics. Very often, we also encounter such systems which have no model and and one has to study such a system from a single observable. So to overcome this lacking feature of the linear analysis and to analyze complex system, one needs resort nonlinear dynamics approach especially chaos theory. The most direct link between chaos theory and the real world is the analysis of the time series from real systems in terms of nonlinear dynamics~\cite{book:Kantz}.  Based on chaos theory, nonlinear tools have been developed to characterize such complicated data. The scope of these methods ranges from invariants such as Lyapunov exponents, correlation dimensions, etc.~\cite{physrevlett:grassberger,PhysicaD:wolf,physicaD:rosenstein}, which yield an accurate description of the structure of the system to statistical tools which allow for classification and diagnosis even in situations where determinism is not obvious.

We have used both types of analysis techniques (linear and nonlinear) to analyze the time series data obtained from our experiments, because some features that could not be revealed by one type of technique was possible by the other.  A detailed  discussion on  the data analysis techniques has been presented in Section~\ref{subsection:analysis} of Section~\ref{section:experiment}.

\section{Experiment, diagnostics, and data analysis procedures}
\label{section:experiment}
\subsection{ Experiment}
\label{subsection:experiment}

\begin{figure} [ht]
 \centering
\includegraphics[width=3in,height=2.75in]{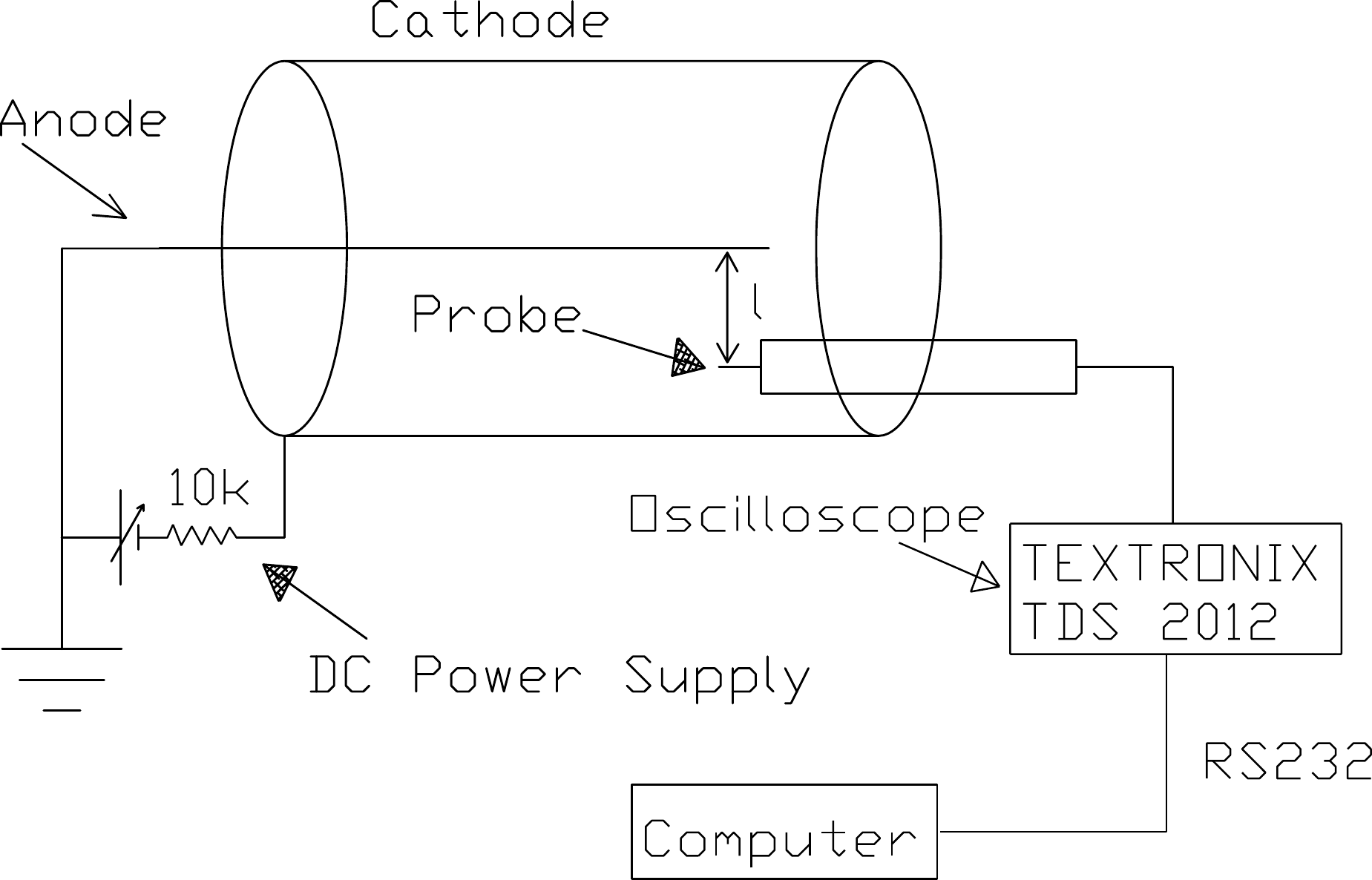}
\caption{Schematic diagram of the cylindrical electrode system of the glow discharge plasma. The probe is placed at a distance $l \approx12.5$ mm from the anode. }
\label{fig:setup}
\end{figure}

\begin{figure*} [ht]
\centering
\includegraphics{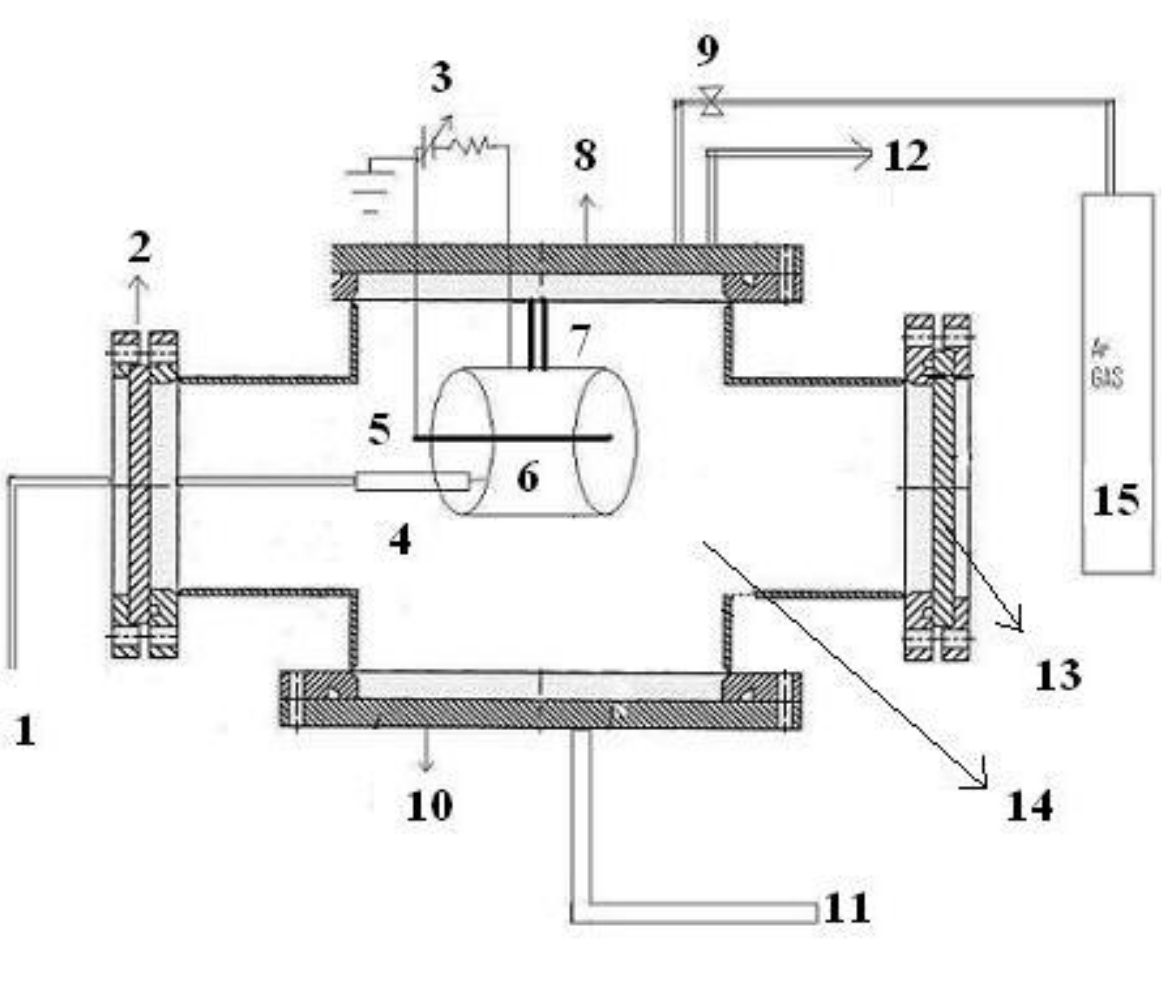}
\caption{ Schematic diagram of the whole experimental setup. 1. probe connection to oscilloscope, 2. view port, 3. power supply, 4. Langmuir probe, 5. anode, 6. cathode, 7. suspension of electrode system, 8. upper flange, 9. needle valve, 10. bottom flange, 11. SS304 bellow, connection to the rotary pump, 12. connection to pirani gauge, 13. toughened glass inside the view port, 14. vacuum chamber, and 15. argon gas cylinder; NG: noise generator, and SG: signal generator.  }
\label{fig:schematic}
\end{figure*}

\begin{figure*} [ht]
\centering
\includegraphics{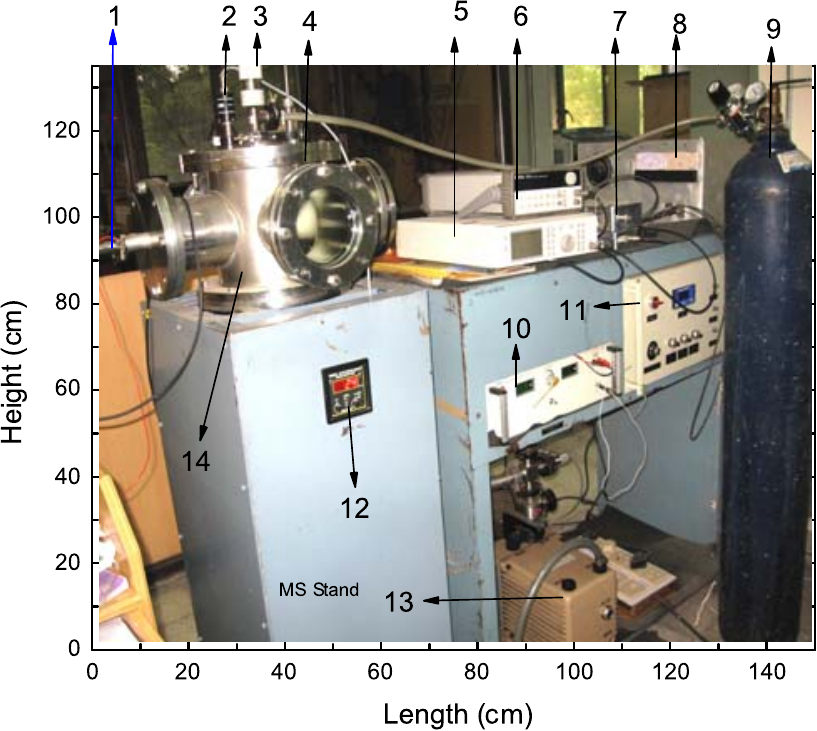}
\caption{ The experimental setup: 1. Langmuir probe, 2. pirani gauge, 3. gas inlet valve, 4. view port, 5. signal generator, 6. noise generator, 7. signal and noise amplifiers, 8. amplifier power supply, 9. argon gas cylinder, 10. discharge power supply, 11. Langmuir probe supply, 12. pressure meter, and 13. rotary pump.  }
\label{fig:wholesetup}
\end{figure*}

The experiments were carried out in a glow discharge argon plasma produced by a dc discharge in a cylindrical hollow cathode electrode system with a typical density and temperature $\approx10^7$ and $2-6$ eV respectively. The schematic diagram of the  electrode system is shown in Fig.~\ref{fig:setup} and its detail is presented in Subsection~\ref{subsubsection:electrode}. The electrode assembly was housed inside a vacuum chamber as shown in Fig.~\ref{fig:schematic}. The neutral pressure inside the vacuum chamber was controlled by a needle valve and the range of the gas pressure in these experiments were between  0.001 to a few  mbar.  The discharge voltage ($0-999$ V) was applied between the cathode and the anode keeping the anode grounded. A noise  (HP33120A) and a signal generator (Fluke PM5138A) were coupled with the power supply for the noise induced experiments. The main diagnostics was the Langmuir probe (described in Subsection~\ref{subsubsection:probe}) used to obtain the floating potential fluctuations. In these experiments the discharge voltage (DV) and pressure were the control parameters.
 Fig.~\ref{fig:wholesetup} is the laboratory view of the whole experimental system and the details of the various parts are presented below.

 \begin{figure} [ht]
\centering
\includegraphics[width=3.5in]{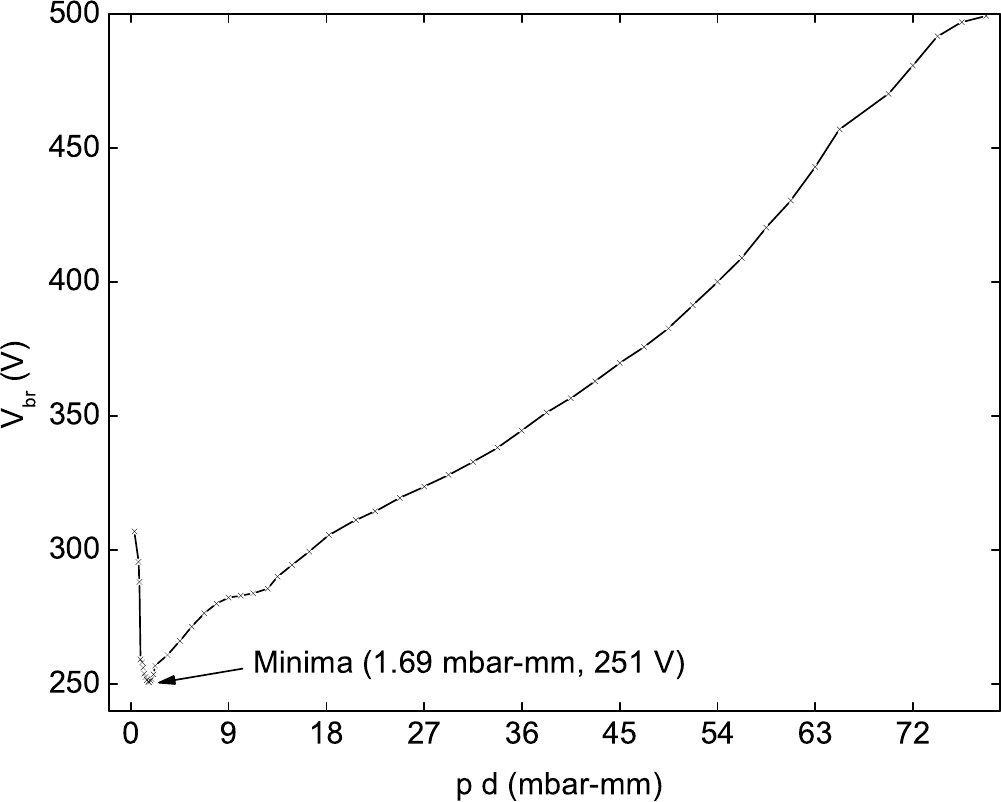}
\caption{$V_{br}$ vs $pd$ (Paschen curve) for our experimental system. The minimum occurs at (1.69 mbar-mm, 251 V).}
\label{fig:paschen}
\end{figure}

By varying the neutral pressure ($p$), the breakdown of the gas occurs at different DV  as shown in Fig~\ref{fig:paschen}. The breakdown voltage ($V_{br}$) initially decreases with increase in $pd$, (where, d is the radius of the cathode)  and then begins to increase with $pd$ after going through a minimum value resembling  a typical Paschen curve~\cite{chaos:nurujjaman,book:von}.  This curve classifies the operating regions of $p$ and DV for the experiments presented in this thesis. When the discharge has been operated for pressures greater than the Paschen minimum, the floating potential fluctuations show chaos to an ordered state transition and has been discussed in Section~\ref{Section:chaos}  and for  less than the Paschen minimum, fluctuations show self organized criticality behavior  and  has been discussed in Section~\ref{section:soc}.  Chaos to an ordered state transition is related to the formation of an anode glow or an anode spot on the anode~\cite{chaos:nurujjaman,JphysD:song, ppcf:sanduloviciu}. Evolution of the glow was monitored by a CCD camera. The ordered state, i.e., when the floating potential is in a steady state which is basically an excitable state is very useful to perform noise invoked resonance experiments and has been presented in Section~\ref{section:resonance}.

 The detail descriptions of the  different parts of the experimental setup have been given in the following subsections.

\subsubsection{Electrode system}
\label{subsubsection:electrode}
\begin{figure}[ht]
\centering
\includegraphics[width=8.5cm]{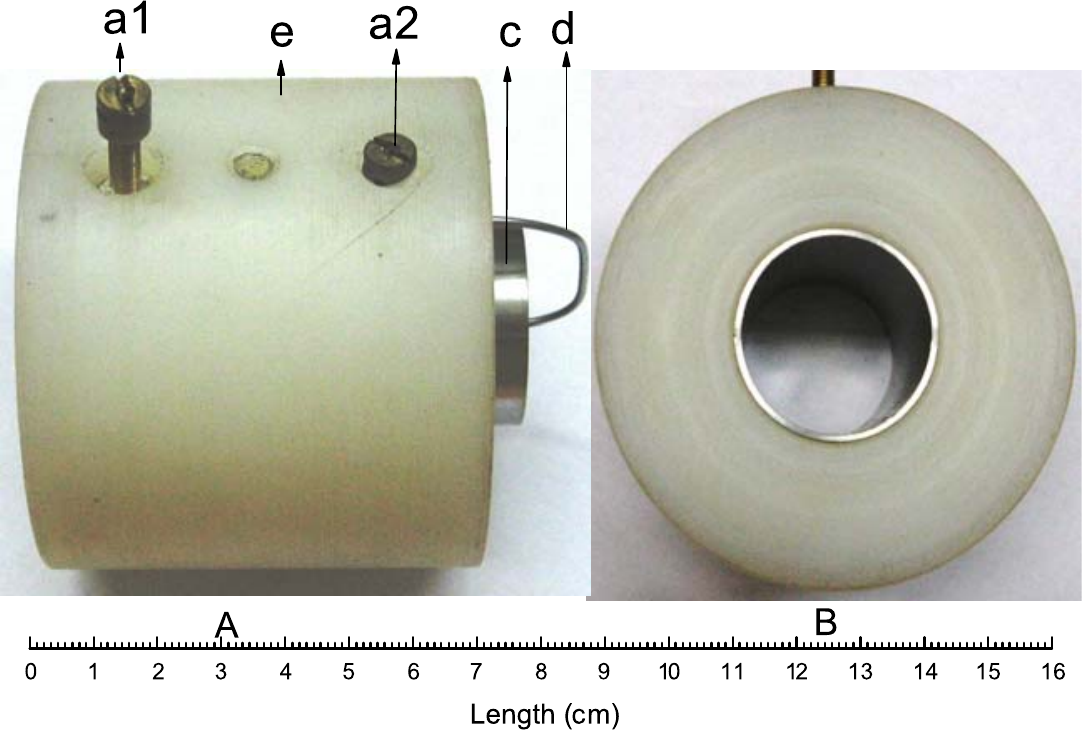}
\caption{Experimental electrode system: A. side view and B. front view of the surface; a1 and a2- connection to the anode and cathode, c. cathode, d. anode, and e. nylon coat on the electrode. }
\label{fig:electrode}
\end{figure}

The experiment has been carried out in a hollow cathode cylindrical electrode system  as shown in Fig~\ref{fig:electrode} and schematic diagram in Fig.~\ref{fig:setup}. Fig~\ref{fig:electrode} (A) and ~\ref{fig:electrode} (B) are the side and front view of the electrode system respectively. The outer stainless steel (SS) cylinder [Fig~\ref{fig:electrode}(c)] is the cathode and the inner SS cylindrical rod placed concentrically with the cathode [Fig~\ref{fig:electrode}(d)] is the anode. Cathode was kept covered with the nylon [Fig~\ref{fig:electrode}(e)]. Electrical connections have been given to the electrodes by two connectors  shown in Figs~\ref{fig:electrode}(a1) and (a2). Almost all the experiments have been carried out using a cathode of diameter 45 mm.

\subsubsection{Vacuum Chamber}
\label{subsubsection:chamber}


SS304 vacuum chamber consists of three parts, bottom flange, main cylindrical body and upper flange [Fig.~\ref{fig:schematic}].  The diameter of the upper and bottom flanges are 300 mm each and height of 300 mm. We have used two upper flanges. It consists of two NW16 coupling through which a gas inlet valve and a Pirani gauge have been connected and few NW25 coupling for inserting electrode system, probes etc.  The bottom flange has a pumping port. The whole chamber is installed on an MS stand [Fig.~\ref{fig:wholesetup}]. There are four side ports each of diameter 200 mm. These ports may be covered using toughened glass or blank flanges or  other suitable flanges fitted with diagnostics.

\subsubsection{Power supply}
\label{subsubsection:supply}
Discharge voltage has been supplied to the electrode system by a dc power supply [Figs.~\ref{fig:schematic} and ~\ref{fig:wholesetup}]. It is a variable power supply of range 0-999 V and maximum output current is 1 amp.  Over current protection and short circuit protection  are available.

\subsubsection{Pumping System}
\label{subsubsection:pump}
The pumping system is a mechanical rotary pump provided by Hind High Vacuum, Bangalore. This pump  has been mounted on the bed of a MS table [Fig.~\ref{fig:wholesetup}]. The vacuum chamber is connected to the rotary pump through an SS bellow followed by a butterfly valve. During pumping the butterfly valve is kept open, and closed when the system is not under operation. The Pirani gauge head is connected on the top flange of the chamber to measure the pressure.

\subsection{Diagnostics}
\label{subsection:diagnostics}
\subsubsection{Langmuir probe}
 \label{subsubsection:probe}

\begin{figure}[ht]
\centering
\includegraphics[width=8.5cm]{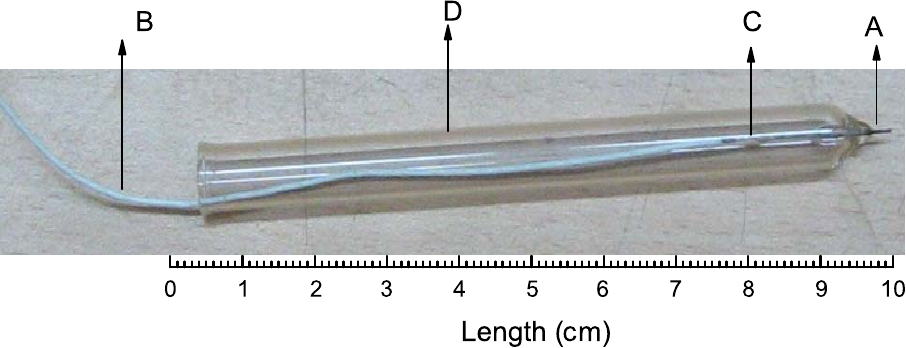}
\caption{Typical glass covered Langmuir probe. A. Tip of the probe (tungsten wire), B. teflon coated wire, C. soldering point of the probe with teflon coated wire, and D. glass cover.}
\label{fig:langmuir}
\end{figure}

 A Langmuir probe is a conductor introduced in the plasma  to measure density, floating potential fluctuations, etc. It consists of a cylindrical tungsten wire of diameter 0.5 mm  of   length $\approx20$ mm [Fig.~\ref{fig:langmuir}]. A teflon coated stainless steel wire has been soldered to the tungsten wire, and it has been fitted inside a glass tube as shown in Fig.~\ref{fig:langmuir}. The exposed tungsten wire of 2 mm length, outside the glass tube is the effective measuring area of the probe. It has been inserted in the plasma through a side port by NW25 coupling and has the facility of axial movement. When it is inserted in the plasma, the electrons and ions fall on it. The imbalance between the ion and electron current, due to the fact that the electrons have a higher mobility between the two, results in an accumulation of the electrons on the surface of the probe, setting up an electric field.  This electric field repels the electrons and attracts ions towards the probe so that the net current is zero and hence the surface acquires a negative potential, which is called floating potential ($\phi_f$).  In the present experiments, the floating potential and its fluctuations have been measured using a tektronix digital oscilloscope.

\subsubsection{CCD Camera}
\begin{figure}[ht]
\centering
\includegraphics[width=9cm]{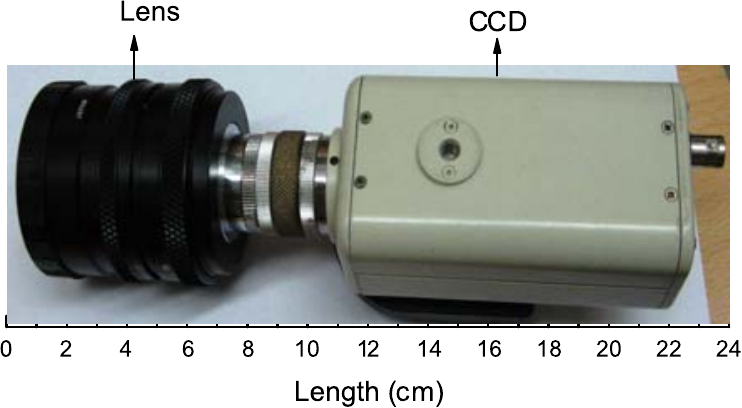}
\caption{Black and white CCD camera used in the experiment. specifications have been given in Table~\ref{table:ccd}. }
\label{fig:ccd}
\end{figure}

Ionization and excitation are both important in glow discharge plasmas. The excitation, followed by de-excitations with the emission  of radiation, is responsible for the appearance of glow. Depending upon different plasma parameters, the size of the glow changes and hence the evolution of it, is an useful observable for the  study of the nonlinear behavior of the glow discharge plasma. The evolution of the glow has been monitored using a black \& white CCD camera shown in Fig~\ref{fig:ccd}. Size of the anode glow has been estimated from the pixel size of the digital image taken by this camera.  Detailed specifications of the camera have been given in Table~\ref{table:ccd}.

\begin{table}[h]
\begin{center} {\footnotesize
\begin{tabular}{|l|l|}
\hline
\multicolumn{2}{|c|}{ CCD Camera} \\
\hline
Image sensor: &1/3" B/W CCC \\
Picture element: & EIA: 768X494, CCIR: 752X582 (HXV)\\
Sync.system: & Internal\\
S/N Ration: & More than 48 dB \\
Horizontal Resolution: & 600 TV lines \\
Min Illumination: & min 0,1 Lux (F 1.2) \\
Auto Gain control: & Built in. \\
Lens mount: & C/CS \\
Video output: & 1 V p.p./75ohm \\
Frame Rate: & 30 FPS-NTFC\\
\hline
\end{tabular} }
\end{center}
\caption{ Specifications of the CCD camera used in the experiments.}
\label{table:ccd}
\end{table}

\subsection{ Data analysis techniques}
\label{subsection:analysis}
The analysis of the floating potential fluctuations has been carried out using the linear and nonlinear time series analysis. In  linear analysis, power spectrum, autocorrelation, probability density function, etc.,  and in nonlinear analysis Lyapunov exponent, correlation dimension, Hurst exponent, etc., have been used. Some statistical tools like `normal variance' and `absolute mean difference' have also been used to characterize coherence and stochastic resonances.  The analysis has been carried out in time as well as frequency domain. Some basics of time series analysis techniques and the procedures followed for computation have been presented in the following subsections.

\subsubsection{Power spectral method}
\label{subsubsection:power}
A time series which has a certain periodicity may be represented as a superposition of periodic components of sines and cosines. The determination of their relative strengths is basically known as spectral or Fourier analysis. When the time series is periodic, the spectrum may be expressed as a linear combination of oscillations whose frequencies are integer multiples of the basic frequency which is the Fourier series. In order to find the characteristic modes, and the dominant mode  present in the floating potential fluctuations, power spectral method has been used. It can also be used to explore other characteristic features present in a time series, such as, self organized criticality behavior, wherein, power spectra shows a  power law behavior and it has been discussed in chapter~\ref{chapter:soc}.  Power spectrum has been calculated using fast Fourier transform method using MATLAB, which has been given below.\\
------------------------------\\
close all   \\
clear all    \\
clc              \\
load data (v)   \\
(power\_v,frequency\_v) = power(v,sampling\_frequency)\\
\% power\_v and frequency\_v are the power and frequency\\
\% of the signal v which has taken at a sampling rate sampling\_frequency\\
plot(frequency\_v,power\_v )\\

\textbf{function [pyy,f]=power\_sig(v,smp\_freq) } \\
\%\\
\% [pyy, f]=power\_sig(v, smp\_freq) calculates power spectrum\\
\% of the signal `v', sampled at frequency `smp\_freq'. \%`pyy' is the output \\
\% power and `f' is the frequency.  \\
\% Written Md. Nurujjaman 2006. \\

\%\\
sig=v;  \\
z=smp\_freq;    \\
lnth\_sig=length(sig);   \\
q=floor(log(lnth\_sig)./log(2));  \\
y = fft(sig,$2^q$);      \\
py = y.* conj(y) /($2^q$);   \\
f = z*(0:($2^q$)/2)/($2^q$);    \\
pyy=py(1:($2^q$)/2+1);      \\

\subsubsection{Autocorrelation}
\label{subsubsection:autocorrelation}
The autocorrelation time of a fluctuating signal is a measure of the temporal coherence and is obtained from the autocorrelation of the signal. For a time series of length n, $X=[X_i,~i=1,2,...n]$, the autocorrelation function (ACF) can be written as ~\cite{PhysicaA:Davide}
\begin{equation}
C(\tau)=\frac{\frac{1}{n-\tau}\sum^{n-\tau}_{j=1}(X_{j+\tau}-\overline{X})(X_j-\overline{X})}{\frac{1}{n}\sum^n_{j=i}(X_j-\overline{X})^2}
\end{equation}
where $\overline{X}$, and $\tau$ are the mean, and time lag of the time series respectively.
If there is long-range time dependence in the signal, then the algebraic decay of the ACF can be written as ~\cite{PRE:GRangarajan2}

\begin{equation}
C(\tau)\sim \tau^{-\alpha}
\end{equation}
for large $ \tau$, where $0<\alpha<1$.

Presence of power law decay of the tail of the ACF indicates the presence of long range correlation in the time series that is one of the characteristic behavior of the self organized criticality [Section~\ref{section:soc}]. Autocorrelation also gives the good estimate of the  time delay of embedding in the phase space that will be discussed in the next sections. It has been calculated using MATLAB in-built function `\textbf{xcorr}'.

\subsubsection{Phase space plot}
\label{subsubsection:phase}
In order to observe nonlinear structures present in a time series and to estimate the correlation dimension, Lyapunov exponent, etc.[subsections~\ref{subsubsection:correlation} and \ref{subsubsection:lyapunov}], the time series has to be described in a suitable phase space. The phase space description  provides a powerful tool for describing the behavior  of a time series in a geometric form.
As in the real experiments, the observable is the sequence of scalar measurements, and one  need to reconstruct phase space vectors from these scalar observable. This can be done by method of delay.

From a time series $(x_1,x_2,..., x_N)$, where N is the total number of points, the m dimensional vector in the phase space can be constructed by delay embedding~\cite{physicaD:rosenstein,physreports:schreiber,physicaD:judd,physrevlett:packard,sringer:takens}
\begin{equation}
X_i=[x_i, x_{i+1}, ..., x_{(i+(m-1)\tau)}]
\label{eqn:embed}
\end{equation}

The time difference ($\tau$)  between two consecutive components of the delay vectors [Eqn.~\ref{eqn:embed}] is referred to as the \emph{lag} or \emph{ time delay}.  The minimum embedding dimension (m)  should be greater than $2D_F$ for good reconstruction of delay vector, where, $D_F$ is the box counting dimension of the attractor. A good estimate of  $\tau$ is more difficult to obtain. As in the mathematical sense, embeddings with the same m but with different $\tau$ are equivalent for noise free data, but in the real case the reconstructed dynamics depends upon $\tau$. There are several rules to choose a time delay. A good time delay may be chosen as $\tau=1-1/e$, of the autocorrelation of the initial value~\cite{book:Kantz}.  Then the reconstructed trajectory of the actual dynamics can be written as
$X=(X_1;X_2;X_3; ...; X_M)$, where $M=N-(m-1)\tau$. This trajectory is useful to estimate the correlation dimension and Lyapunov exponents that have been discussed in the next subsections.
\subsubsection{Correlation dimension}
\label{subsubsection:correlation}
The correlation dimension of a time series is defined as the dimensionality of the space occupied by the points of that time series. It can be estimated from the correlation integral of the reconstructed trajectory in phase space [subsection~\ref{subsection:phase}] of the time series. The correlation integral can be computed using following equation~\cite{physrevlett:grassberger,physicaD:rosenstein,book:Kantz,physrevA:grassberger}
\begin{equation}
C(r)=\frac{2}{N(N-1)}\sum_{i=1}^N\sum_{j=i+1}^N\Theta(r-|X_i-X_j|)
\end{equation}
where, $r$ is scale length, and $\Theta$ is the Heaviside step function.    $X_i$, $X_j$ will be obtained using Equation~\ref{eqn:embed}.

The scaling of the function $C(r,m)$ can be written as
\begin{equation}
C(r,m)=r^D
\end{equation}
where $D$ is the Correlation dimension defined as
\begin{equation}
D=\lim_{r \rightarrow 0}\lim_{N\rightarrow \infty}\frac{\partial C(r,m)}{\partial \ln r}
\end{equation}
and can be obtained from the slope of $\ln C(r)$ vs $ \ln r$ plot.
We have used the algorithm written by Rosenstein in C for computing correlation dimensions.

\subsubsection{Lyapunov exponent}
\label{subsubsection:lyapunov}
Chaotic dynamical systems are sensitive to initial conditions, and exhibit an exponential divergence in the phase space. The divergence can be quantified by an exponent called Lyapunov exponent. There are several methods to estimate Lyapunov exponent. Here  Rosenstein method has been used, because this method is very useful for short data length.

  If we consider two points on two nearby trajectories of a chaotic attractor, in the phase space, assuming the distance between them to be $d(0)$, and after time t, if the distance between the two trajectories becomes $d(t)$, then the divergence (separation after time t) can be written as~\cite{physicaD:rosenstein}

\begin{equation}
d(t)=d(0)e^{\lambda_L t}
\end{equation}
where $\lambda_L$ is the largest Lyapunov exponent. Since a practical time series is basically a scalar measurement, the largest $\lambda_L$ can be estimated from the reconstructed trajectories in the phase space.   Now if $X_j$ and $X_{\hat{j}}$ are the $j^{th}$ pair of the nearest neighbor on two nearby reconstructed trajectories of that time series in the phase space and distance between them is $d_j(0)=min_{X_j}||X_j-X_{\hat{j}}||$, then the separation after time $i\Delta t$, where $\Delta t$ is the sampling time, can be written as
\begin{equation}
d_j(t)=d_j(0)e^{\lambda_L(i\Delta t)}
\end{equation}
where $d_j(0)$ is the initial separation. The above equation can be written as
\begin{equation}
\ln d_j(t)=\ln d_j(0) +\lambda_L(i\Delta t)
\end{equation}
and the largest $\lambda_L$ can be calculated from the slope of the average line defined by
\begin{equation}
<\ln d_j(t)>=<\ln d_j(0)> +\lambda_L(i\Delta t)
\end{equation}

$\lambda_L$ quantifies the chaotic nature of the fluctuation, and  +Ve $\lambda_L$ indicates the presence of chaos. It has been calculated using the algorithm developed by Rosenstein in C.

 \subsubsection{Surrogate method}
The interest in detecting determinism or stochasticity  in time series arose out of the need to discriminate a chaotic time series from a random one.  To do this one has to first generate surrogate time series from the original time series using some specific rules which is termed as surrogate method. Once the property under investigation (for instance nonlinearity,
determinism, and so on) is defined, a method is sought to generate the surrogate data that
preserves a specific characteristic of the original time series. A discriminant factor is then
used to compare the original time series with the time series generated by a suitably
chosen surrogate method. If there are significant difference between the discriminating factors obtained from the two time series, then one can say that the original is not random.

Several algorithms have been proposed to generate the surrogate data in the literature.
The most commonly used methods are: Random Shuffled, phase shuffled and Amplitude Adjusted Fourier shuffled methods~\cite{physicaD:theiler,ieee:small,Bifur:Nakamura}. We have used  phase shuffled method for our experimental data. The correlation dimension has been used as the discriminating factor.  The phases of the original time series have been randomized by Phase Shuffled method, by shuffling the fourier phases~\cite{physicaD:theiler,Bifur:Nakamura,chaos:dori}. This method preserves the power spectrum (linear structure), but the nonlinear structures are destroyed~\cite{Bifur:Nakamura}.  Now if the original time series has some kinds of nonlinearity, the correlation dimension estimated from the both times series will differ significantly and hence useful to detect determinism present in the time series. The algorithm written by Michael Small using MATLAB has been used to generate the surrogate data and the codes may be found in his web page.

\subsubsection{Hurst exponent}
The rescaled-range statistics $(R/S)$ method was proposed by Hurst and well established by Mandelbrot, and Wallis~\cite{POP:carreras}. For the time series defined above [subsection~\ref{subsection:phase}], the $R/S$ is defined as~\cite{POP:carreras} the ratio of the maximal range of the integrated signal normalized to the standard deviation:
\begin{equation}
\frac{R(n)}{S(n)}=\frac{max(0,W_1,W_2,...,W_n)-min(0,W_1,W_2,...,W_n)}{\sqrt{S^2(n)}}
\end{equation}
Here $W_k= x_1+x_2+x_3+...+x_k-k\overline{X(n)}$, where $\overline{X}$, $S^{2}(n)$, and n are respectively the mean, variance, and time lag of the signal. The expected value of $R/S$ scales like $cn^H$ as $n\rightarrow \infty$, where H is called the Hurst exponent. For random data H=0.5, while $H>0.5$  for the data with long range correlations. $H<0.5$ indicates the presence of long-range anti-correlation in the data.

Presence of long range correlation in a time series  can be estimated using Hurst exponent. The algorithm for estimating Hurst exponent was written in MATLAB.
\\
---------------------\\
 close all \\
clear all\\
clc\\
load data(v)\\
(ln\_n,ln\_rs1)=hurst\_exp(v) ;\\
 plot(ln\_lag,ln\_rs);\\
 \% Slope of the fit will give Hurst exponent\\

\textbf{function [ln\_n,ln\_rs1]=hurst\_exp(v1)}\\
\%\\
\%  [ln\_n,ln\_rs1]=hurst\_exp(v1) Calculates HURST exponent\\
\% of a signal v1.
\%ln\_n is the log of lag  and ln\_rs1 is the log of (R/S).\\
\% v1 is the time series.\\
\%  exponent of a vector V.\\
\% Last Modified 02/11/2007\\
\% MD. NURUJJAMAN\\
\%=======================================\\

v=v1;\\
ll=length(v);\\
nl=floor(ll/100);  \% One can define the block smallest data length using this.\\
i=1;\\
while i$<$=nl\\
    v1=[ ];\\
    j=1;\\
    ll\_f=floor(ll./i);\\
   \%\\
     while  j$<$=i\\
        v1(j)=(max(cumsum(v((ll\_f*(j-1)+1):(ll\_f*j))...\\  -nanmean(v((ll\_f*(j-1)+1):(ll\_f*j)))))...\\
            -min(cumsum(v((ll\_f*(j-1)+1):(ll\_f*j))...\\ -nanmean(v((ll\_f*(j-1)+1):(ll\_f*j))))))...\\
            /nanstd(v((ll\_f*(j-1)+1):(ll\_f*j)));\\
        j=j+1;\\
     end\\
   \%\\
\\
    rs(i)=nanmean(v1);\\
    ln\_n(nl+1-i)=log(ll/i);\\
    ln\_rs1(nl+1-i)=log(rs(i));\\
    i=i+1;\\
end\\
\subsubsection{Normal variance}
Normal variance (NV) has been used to quantify the regularity in the signal recovered by introducing noise in the plasma. NV~\cite{prl:pikovsky} is defined as $NV=std(t_p)/mean(t_p)$ where, $std(t_p)$ the standard deviation of the peak to peak distance ($t_p$) of the recovered signal. If the distribution of the peaks are regular NV is small and hence it is the estimation of the coherency present in the system. NV has been calculated using  algorithm written in  MATLAB.

\textbf{function stat=NV(v, delta, smpleT)}\\
 \%\\
\% NV estimates Normalized variance of time series v.\\
\% NV is defined as NV=standard deviation(v)/mean (v)\\
\% delta is the height of the maximum from its preceded value.\\
\% smpleT is the sampling time of the signal.\\
\%\\
 (maxtab, mintab)=peakdet(v, delta);\\
\%\\
\% max\_index = maxtab(:,1); \% index of the peaks\\
 st=smpleT;\\
 max\_index1=max\_index(2:end);\\
 max\_index2=max\_index(1:end-1);\\
 Max\_Diff=max\_index1-max\_index2;\\
 P2Ptm=Max\_Diff.*st;\\
stat=std(P2Ptm)/mean(P2Ptm);\\
\%\\
function [maxtab, mintab]=peakdet(v, delta)\\
\%\\
\%PEAKDET Detect peaks in a vector\\
\%        [MAXTAB, MINTAB] = PEAKDET(V, DELTA) finds the local\\
\%        maxima and minima (``peaks") in the vector V.\\
\%        A point is considered a maximum peak if it has the maximal\\
\%        value, and was preceded (to the left) by a value lower by\\
\%        DELTA. MAXTAB and MINTAB consists of two columns. Column 1\\
\%        contains indices in V, and column 2 the found values.\\
\\
\% Eli Billauer, 3.4.05 (Explicitly not copyrighted).\\
\% This function is released to the public domain; Any use is allowed.\\
\\
maxtab = [];\\
mintab = [];\\
\\
v = v(:); \% Just in case this wasn't a proper vector\\
\\
if (length(delta(:)))$>$1\\
  error(`Input argument DELTA must be a scalar');\\
end\\
\\
if delta $<=$ 0\\
  error(`Input argument DELTA must be positive');\\
end\\
\\
mn = Inf; mx = -Inf;\\
mnpos = NaN; mxpos = NaN;\\
\\
lookformax = 1;\\
\\
for i=1:length(v)\\
  this = v(i);\\
  if this $>$ mx, mx = this; mxpos = i; end\\
  if this $<$ mn, mn = this; mnpos = i; end\\
\\
  if lookformax\\
    if this $<$ mx-delta\\
      maxtab = [maxtab ; mxpos mx];\\
      mn = this; mnpos = i;\\
      lookformax = 0;\\
    end\\
  else\\
    if this $>$ mn+delta\\
      mintab = [mintab ; mnpos mn];\\
      mx = this; mxpos = i;\\
      lookformax = 1;\\
    end\\
  end\\
end\\

\subsubsection{AMD}
Absolute mean difference (AMD)~\cite{pre:nurujjaman} is the statistical tool, proposed to quantify the stochastic resonance in a plasma subjected to noise and a periodic signal. AMD is defined as $AMD=abs(mean(t_p/\delta-1))$, and gives  AMD gives the degree of mimicking the output to the input subthreshold signal.
Usually, regularity in the stochastic resonance is quantified
by calculating cross-correlation ($Co = | < [(x1- < x1 >
)(x2- < x2 >)] > |$) between the output and input signal. But in case of plasma it is not suitable,
because there is always a lag between periodic signal that is applied to the plasma and
the output. This lag also varies with time because the plasma
conditions keep fluctuating over time. Therefore,
cross-correlation is not the right quantity to be estimated. So we have proposed a statistical tool AMD which will be independent of lag and is estimated as follows:
\begin{enumerate}
\item First calculate the mean inter-peak distance ($\delta$) of the
periodic signal.
\item Calculate the inter-peak distances ($t_p$) of the output
signal.
\item Calculate the ($(t_p -\delta)/\delta$)
\item Take the absolute, i.e.,    $| < ( t_p/\delta - 1) > |$.

\end{enumerate}

Therefore, $AMD = | < ( t_p/\delta - 1) > |$.

The algoritm for AMD is almost identical of Normal variance.

\section{Chaos to order and homoclinic bifurcation in glow discharge plasma}
\label{section:chaos}

\subsection{Introduction}
In the first chapter, we had mentioned that in almost all the chaos related plasma experiments, plasma fluctuations were observed to go  to a chaotic state with external control parameters, like discharge voltage, currents, etc. On the other hand, there has been no experimental evidence of inverse routes, i.e., chaos to order. It has only been seen numerically  in plasma by Tomejiro Yamagishi and Makoto Tsukayama~\cite{jpsj:makato1}. In the present experiment, we have reported chaos to an ordered state transition  for a particular range of plasma parameters given by the Paschen curve which has been discussed in Section~\ref{subsection:experiment} of Section~\ref{section:experiment}.  For those pressure regions, with the initiation of discharge, a bright glow or spot [Fig~\ref{fig:glow}] was observed to form on the anode that was unstable initially and was the source of the chaotic fluctuations . So to understand the dynamics of the anode glow one needs to analyze  the floating potential fluctuations  that has been discussed in next Subsection~\ref{subsection:glow}. Nonlinear analysis of the fluctuation  has been carried by estimating the correlation dimension and the Lyapunov exponent.

\subsection{Anode glow}
\label{subsection:glow}
Anode glow is a bright structure or spot of higher plasma density, producing anode double layer by additional excitation and ionization process and forms around an anode or any point on the anode or on a positively biased probe~\cite{JphysD:song,ppcf:sanduloviciu,psst:stenzel,chaossolitonfractal:agop,epl:popescu,applphysD:Song,Phys.Scr.:Song,Phys.Scr.:Song1,pla:Sanduloviciu,pla:Sanduloviciu1,ppcf:Amemiya}. It appears when the discharge current is too low to sustain the discharge~\cite{JphysD:song,ppcf:sanduloviciu}.  Its size and shape differ depending upon the plasma conditions. It is related to highly nonlinear phenomena involving the physics of sheaths, double layers, ionization, beam and possibly external circuit interactions.

\begin{figure}[ht]
\centering
\includegraphics[width=8.5 cm]{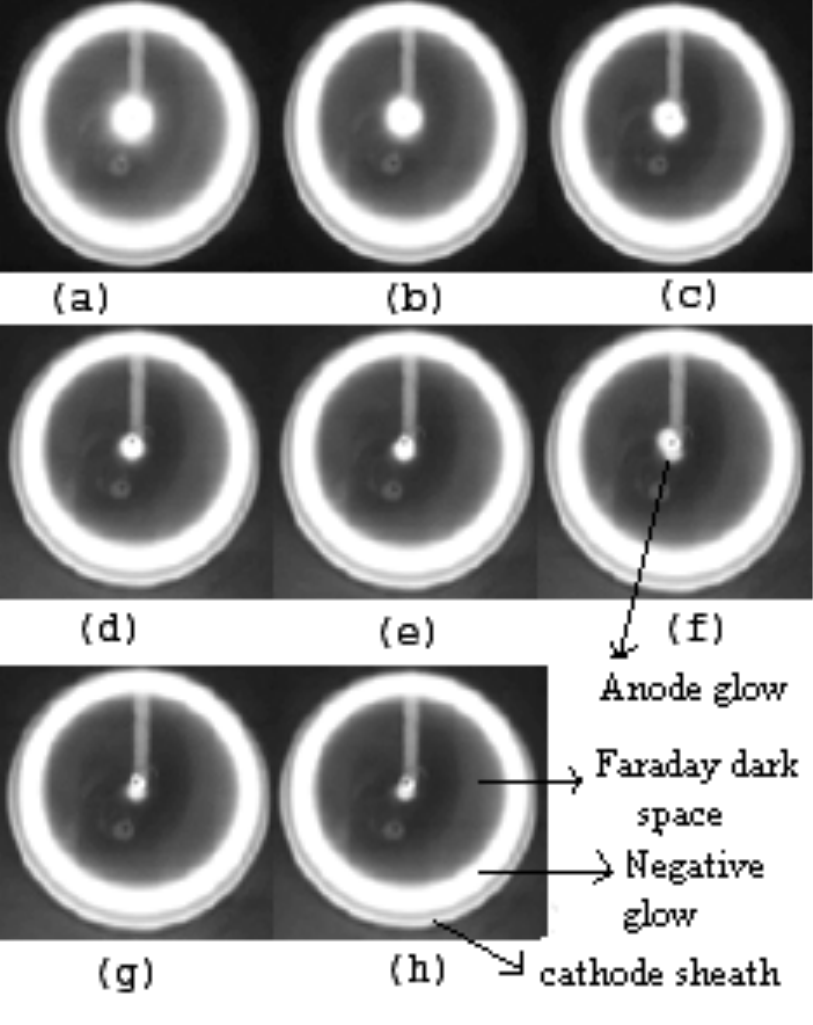}
\caption{Evolution of glow size of the anode glow with increasing DV (a) - (h).}
\label{fig:glow}
\end{figure}

Figs ~\ref{fig:glow}(a) shows that the glow with largest size, appears when the discharge is struck at a typical pressure of 0.95 mbar and its size decreases with increase in the DV until it finally disappears [Figs~\ref{fig:glow}(a)$-$\ref{fig:glow}(h)]. From the CCD image analysis the annular radius of the glow around the anode, was estimated to be $\approx 1.3$ mm at the beginning of the discharge [Fig~\ref{fig:glow}(a)] and reduced to $\approx 0.32$ mm [Fig~\ref{fig:glow}(f)].

 Theoretically one can obtain an expression for the thickness ($\delta$) following Bin Song et al.~\cite{JphysD:song}. The rate of ion production in the sheath is approximately,
 \begin{equation*}
n_{es}\sqrt{\frac{eV_A}{m_e}}\delta\sigma_i(eV_A)N,
\end{equation*}
where, $n_{es}$, $m_e$ and N are the electron density, mass and neutral density respectively.
$V_A$ is the positive voltage created inside the anode, $\delta$ the thickness of the glow, $\sigma_i$ the ionization cross-section depends upon the energy of the electron and at 15.76 eV, $\sigma_i$ is $\approx2\times10^{-18}cm^2$~\cite{jchemphys:rapp}.

The rate of ion loss from the glow is approximately,

 \begin{equation*}
 n_{is}\sqrt{\frac{eV_A}{m_i}},
 \end{equation*}
where, $n_{is}$, $m_i$ are ion mass and density respectively.

 Balancing the ion loss with the ion production, and invoking the condition that $n_{is}\approx n_{es}$, we obtain
  \begin{equation*}
 \sqrt{\frac{m_i}{m_e}}\delta\sigma_i(eV_A)N\approx1
 \end{equation*}
  or
\begin{equation}
\label{eqn:thickness}
\delta\approx3.7\times10^{-6}\frac{kT}{\sigma_{i}P},
\end{equation}
 where k, T, and P are the Boltzmann constant, room temperature in Kelvin scale, and pressure in mbar respectively. The estimated thickness of the anode glow using the Eqn.~\ref{eqn:thickness}, for $P=0.95$ mbar, $T=300$ K, is $\approx0.81$ mm which is  within the range of the thickness estimated from the image ($1.3-0.32$ mm) shown in Fig~\ref{fig:glow}.

\begin{figure}[ht]
\centering
\includegraphics[width=8.5 cm]{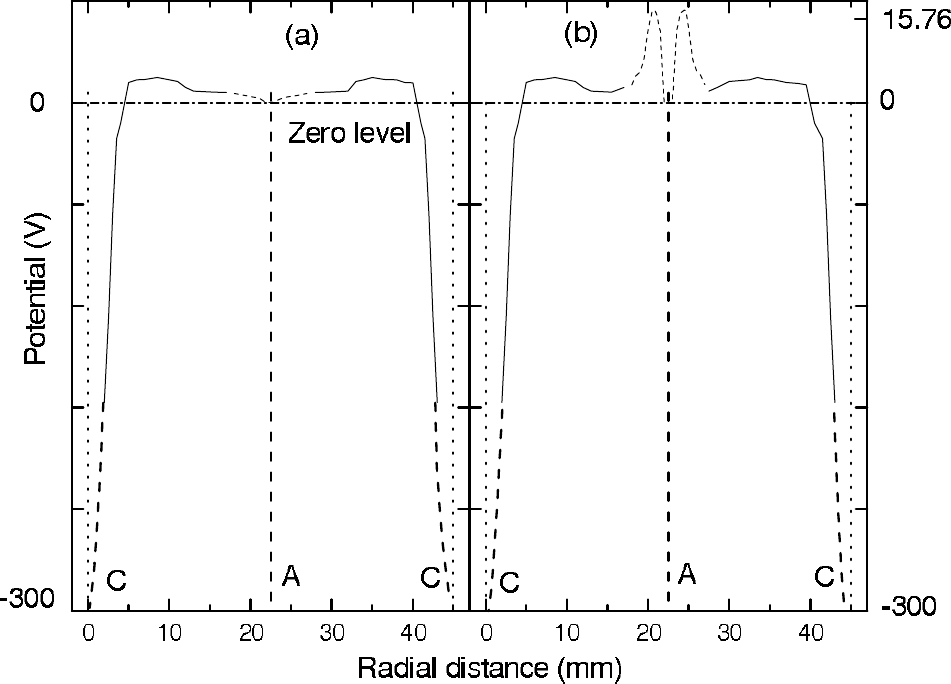}
\caption{Spacial variation of the plasma floating potential: (a) without anode glow; (b) with anode glow. Vertical dotted  and dashed lines are the cathode positions (C) and anode position (A) respectively. Horizontal dash dot line is the zero level. $--$ in the curve near the anode and the cathode are the extrapolation of the potential.}
\label{fig:pot}
\end{figure}

Fig~\ref{fig:pot}(a) and~\ref{fig:pot}(b) show the tentative model of floating potential profile for the present experimental arrangement. It shows that the floating potential increases sharply in the sheath region near the cathode and then decreases slowly up to the anode. As the measurement near the anode and the cathode was not possible, the extrapolated potential profile is shown by dash lines. In the the presence of the anode glow, the potential near the anode get modified. Experimentally it is observed that the anode glow is separated from rest of the plasma by a double layer~\cite{JphysD:song,ppcf:sanduloviciu,JphysD:opresu,pla:sanduloviciu} and the height of the potential of the core of this glow is of the order of the ionization potential~\cite{JphysD:opresu}, which for argon is about 15.76 eV. As the anode is grounded, the modified potential profile in this case, over the profile shown in Fig~\ref{fig:pot}(a), will be as shown in Fig~\ref{fig:pot}(b). As the width of the anode glow decreases with increase in the DVs [Fig~\ref{fig:glow}] the hump shown in Fig~\ref{fig:pot}(b) will also shrink accordingly. This shrinkage happens because as the electron energy increases with the DVs, the necessity of the additional ionization to maintain the discharge decreases~\cite{JphysD:song}. Based on the observation of Valentin Pohoa\c{t}\v{a}, \emph{et al.}~\cite{pre:Valentin}, we feel that the relaxation oscillations are double layer or such coherent potential structures, which are constantly forming and annihilating.

\subsection{Analysis of the floating potential fluctuations}
\label{subsection:fluctuation}
\subsubsection{Chaos to order}
\label{subsubsection:chaos}
\begin{figure}[ht]
\centering
\includegraphics[width=8.5cm]{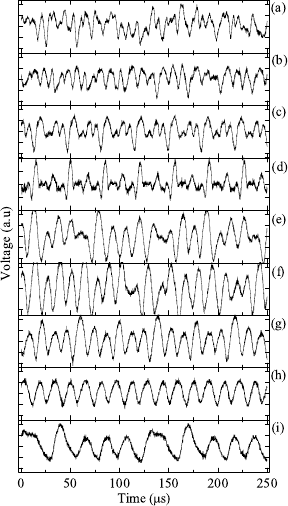}
\caption{Sequential change in the raw signal (normalized) at 1mbar for different voltages: (a) 293 V; (b) 296 V; (c) 298 V; (d) 299 V; (e) 300 V; (f) 305 V; (g) 308 V; (h) 310 V; (i) 312 V. All y-axes range form -1 to 1.}
\label{fig:raw1mb}
\end{figure}

An interesting feature associated with the anode glow was the different types of oscillations in the floating potential at different pressures.
\begin{figure}[ht]
\centering
\includegraphics[width=8.5cm]{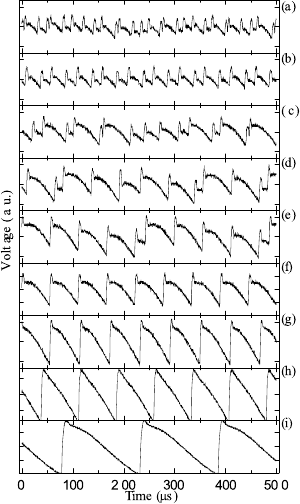}
\caption{Sequential change in the raw signal (normalized) at 0.89 mbar for different voltages: (a) 288 V; (b) 291 V; (c) 295 V; (d) 301 V; (e) 304 V; (f) 307 V; (g) 327 V; (h) 385 V; (i) 466 V. All y-axes range form -1 to 1.}
\label{fig:raw0.89mb}
\end{figure}
We carried out a detailed analysis of the fluctuations for three typical pressures which are presented here. At about 0.89 mbar ($pd\approx 20.02$ mbar-mm), the discharge was initiated at $\approx 288$ V and an anode glow was observed similar to Fig~\ref{fig:glow}(a).
\begin{figure}[ht]
\centering
\includegraphics[width=8.5cm]{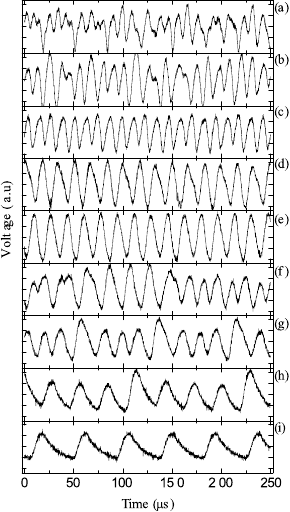}
\caption{Sequential change in the raw signal (normalized) at 0.95 mbar for different voltages:(a) 283 V; (b) 284 V; (c) 286 V; (d) 288 V; (e) 289 V; (f) 290 V; (g) 291 V; (h) 292 V; (i) 293 V. All y-axes range form -1 to 1.}
\label{fig:raw0.95mb}
\end{figure}
Simultaneously irregular relaxation type of oscillations in the floating potential were observed [Fig~\ref{fig:raw0.89mb}(a)]. Increasing the DVs, led to an increase both in the amplitude and the time period of the oscillations [Figs~\ref{fig:raw0.89mb} (b)$-$\ref{fig:raw0.89mb}(i)]. The regularity of the oscillation also increases with the DVs. Around 509 V both the anode glow and the fluctuations disappeared simultaneously and this point is termed as bifurcation point or critical point ($V_H$). At 0.95 mbar ($pd\approx 21.37$ mbar-mm) the fluctuations are observed to be more random [Fig~\ref{fig:raw0.95mb}(a)$-$(i)] than at 0.89 mbar. With increasing DV, the final form of the fluctuations before their disappearance was the relaxation type of oscillation as shown in Figs~\ref{fig:raw0.95mb}(h)$-$\ref{fig:raw0.95mb}(i). Increasing the pressure causes more randomness in the signals as seen in Fig~\ref{fig:raw1mb}(a)$-$(i) at 1.0 mbar ($pd\approx 22.5$ mbar-mm). We observed the relaxation type of oscillations  at DV $\approx313$ V just before the glow and the fluctuation disappears.  So in general, the autonomous dynamics shows transition from oscillatory to steady state behavior.

\textbf{ Linear analysis:}

\begin{figure*}[ht]
\centering
\includegraphics[width=13cm]{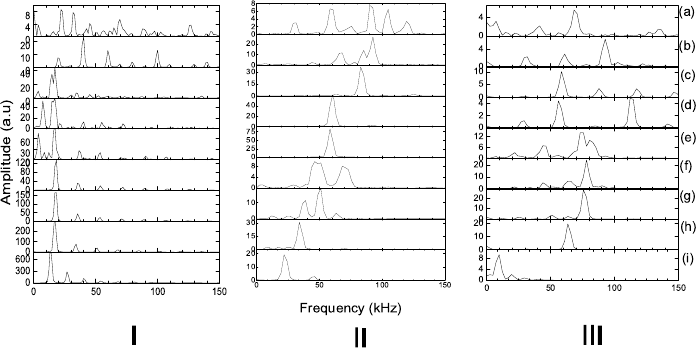}
\caption{Power spectrum of floating potential fluctuations at initial discharge voltages at  filling pressures: (a) 0.89 mbar, (b) 0.95 mbar and (c)1 mbar. }
\label{fig:powspecs}
\end{figure*}

 To understand the nature of the fluctuations we have analyzed them using Fourier techniques.
Figs~\ref{fig:powspecs}\{I\}, \{II\} and \{III\} show the power spectrum~\cite{chaos:nurujjaman,pramana:nurujjaman} calculated for the signals at three different experimental pressures shown in Figs~\ref{fig:raw0.89mb},~\ref{fig:raw0.95mb} and ~\ref{fig:raw1mb} respectively.
\begin{figure}[ht]
\centering
\includegraphics[width=8.5 cm]{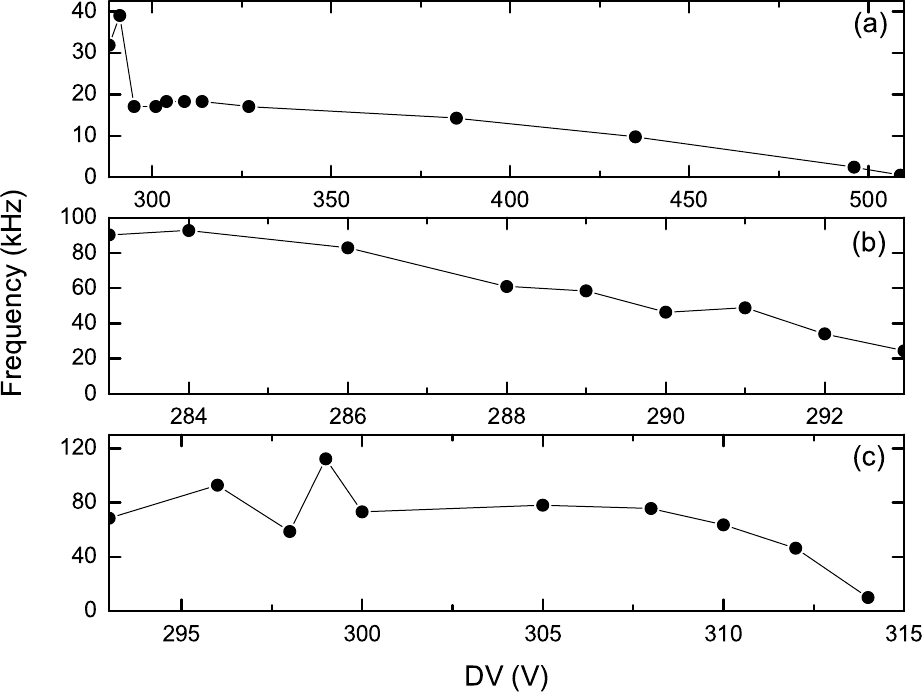}
\caption{Dominant frequency vs DV at the three experimental pressure: (a) 0.89 mbar; (b) 0.95 mbar; (c) 1.0 mbar.}
\label{fig:domifreq}
\end{figure}
A broad range of frequencies ($<$ ion plasma frequency) is observed at the initial stage of the discharge, at pressures 0.89 mbar [Fig~\ref{fig:powspecs}\{I\} (a)], 0.95 mbar [Fig~\ref{fig:powspecs}\{II\} (a)] and 1.0 mbar [Fig~\ref{fig:powspecs}\{III\} (a)$-$(b)]. The figures also show that with increasing DV, number of peaks present in the power spectrum, decreases. At higher DV, i.e., at the end of the sequence of the power spectra we observed one major peak with its harmonic [Figs~\ref{fig:powspecs}\{I\}(f)$-$(i) and III(i)] or two peaks [Figs~\ref{fig:powspecs}\{II\}(i)] just before the fluctuations cease. The power spectra clearly show that the system goes from an irregular to a more or less regular state with increasing DV. Another interesting feature that has been observed for all the three pressures is that the dominant frequency, i.e., the frequency which has maximum power  decreases with increase in the DVs [Figs~\ref{fig:domifreq}(a)$-$(c)].  On the contrary, in many experiments~\cite{pop:Klinger,pop:lee,ppcf:gyergyek}, the dominant frequency is observed to increase with the DV. The basic difference between the present experiment and the experiments performed in Ref~\cite{pop:Klinger,pop:lee,ppcf:gyergyek} was in the electrode configuration and biasing arrangement. In those experiments~\cite{pop:Klinger,pop:lee,ppcf:gyergyek} the system was planar type is which the anode voltage was increased keeping the cathode at ground whereas ours is a cylindrical system in which we increased the negative voltage on the cathode keeping anode at ground. In those systems electrons originating at cathode get lost at the anode. In our case the electrons can undergo multiple reflections and hence increases the ionization length. The fluctuations are probably due to ion acoustic instability driven by electron beam plasma interaction and also from electron and ion trapping in the potential wells. associated with the anode glow [Fig~\ref{fig:pot}].

Ions produced inside the anode glow due to collisions of the accelerated electrons across the hump with the neutrals, makes the glow unstable~\cite{ppcf:sanduloviciu}, which is probably responsible for the relaxation or the random oscillations~\cite{pop:Klinger,pop:lee}. An estimate of the frequency of these instabilities can be obtained from the ion transit time in the plasma~\cite{pop:Klinger}  $\tau(d)=\frac{d}{V_{th,i}}=d/\sqrt\frac{k_bT_i}{m}$, where d is the electrode distance. The estimated ion transit frequency ($\frac{1}{\tau}$) for our experimental system is $\approx19$ kHz which agrees well with the frequency of the relaxation oscillations of the floating potential shown in Figs~\ref{fig:powspecs}\{I\}(f)$-$(i), \{II\}(i) and \{III\}(i). The higher frequency oscillations could be due to ion acoustic instabilities since the conditions are quite conducive to excite these low frequency instabilities. In the next subsection we have presented nonlinear analysis of the fluctuations.

\textbf{Nonlinear analysis:}

The presence of relaxation oscillations have been attributed to the formation of highly nonlinear structures like double layers~\cite{pre:Valentin}. We therefore estimated the correlation dimension ($D_{corr}$) and the +ve Lyapunov exponent ($\lambda_L$) of all the signals. In this experiment, it is observed that the nature of the fluctuations of the potential did not vary for almost the whole day as long as the controlling parameters are kept constant, so the analyzed signals can conveniently be taken to be stationary.
 \begin{figure}[ht]
\center
\includegraphics[width=8.5cm]{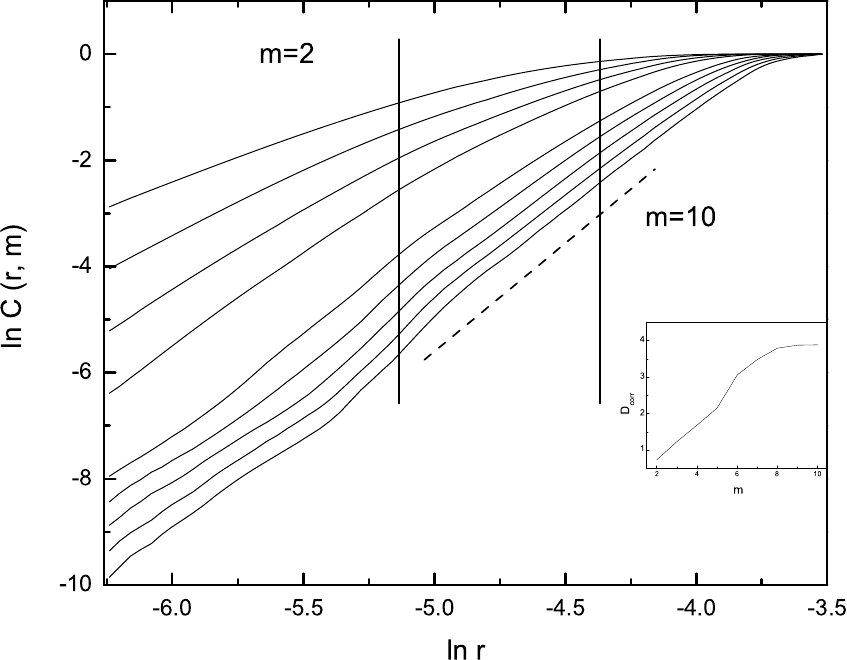}
\caption{ Effect of embedding dimension on correlation sums. The scaling region is shown by two vertical lines. The best fitting is shown by $--$ line.}
\label{fig:corr}
\end{figure}

\begin{figure}[ht]
\center
\includegraphics[width=8.5cm]{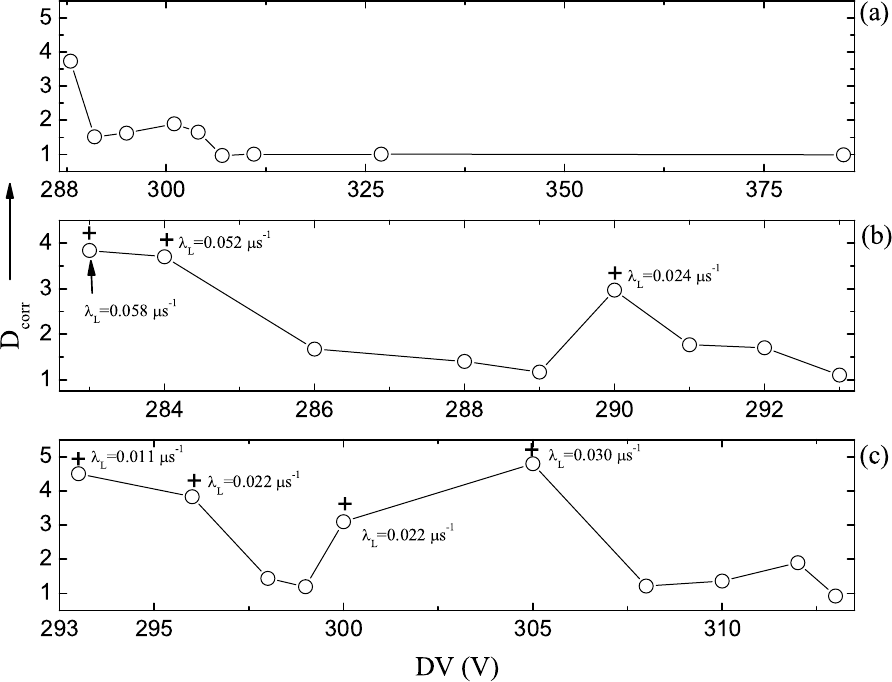}
\caption{$D_{corr}$ vs discharge voltages.  +Ve $\lambda_L$ has been shown by +ve sign. at (a) 0.89 mbar; (b) 0.95 mbar; (c) 1 mbar.}
\label{fig:cordisch}
\end{figure}

The correlation dimension ($D_{corr}$) has been calculated using well-known Grassberger-Procaccia techniques~\cite{physrevlett:grassberger,physrevA:grassberger},  and detail has been discussed in Subsection~\ref{subsection:analysis} of Section~\ref{section:experiment}. A typical plot of $\ln C(r,m)$ vs $\ln r$ has been shown in  Fig~\ref{fig:corr} for the embedding dimension (m), in the range of 2 to 10, from which we have estimated the correlation dimension ($D_{corr}$) for a typical signal (Fig~\ref{fig:raw0.95mb}(a)) at DV $\approx283$ V,  at  0.95 mbar. The scaling region had to be chosen carefully, since for too small scale lengths the correlation sum is heavily distorted by noise and the higher scale lengths are limited by attractor dimensions. From the above plot the correlation sum exhibits a power law behavior within a certain range of $r$ as shown by the vertical dotted line. $\ln C(r,m)$ vs $\ln r$ plots are almost parallel for higher m (i.e. $m=7-10$) and the corresponding best fit has been shown by $--$ line. $D_{corr}$ vs m is also shown in the inset of Fig~\ref{fig:corr}. It shows that $D_{corr}$ becomes constant at higher m and this constant value of $D_{corr}$ is the correlation dimension of that particular signal and for all of our data $D_{corr}\geq3.8$ to begin with. The $D_{corr}$ at 0.89, 0.95 and 1 mbar for different DVs have been shown by open circle($\circ$) in Figs~\ref{fig:cordisch}(a)$-$\ref{fig:cordisch}(c) respectively. It is observed that there is a decreasing tendency  of $D_{corr}$ except for some intermediate values of DV at higher pressures, where it is seen to increase and then decrease again. Since $D_{corr}$ is a measure of the complexity of the system it is likely that the system complexity increases at those intermediate DVs. In our experiment initially the system is in a complex state as $D_{corr}$ for all the three pressures is high and decreases with increase in DVs [Fig~\ref{fig:cordisch}(a)$-$(c)].  $D_{corr}\approx1$ just before the system reaches stable state finally, is the indicator of the periodic state and these periodic nature are also prominent from the raw data and the power spectrum [Figs~\ref{fig:raw0.89mb},~\ref{fig:raw0.95mb}, ~\ref{fig:raw1mb} and Fig~\ref{fig:powspecs}\{I\}$-$\{III\}]. $D_{corr}$ shows that the system stabilizes itself with increase in DVs.

\begin{figure}[ht]
\center
\includegraphics[width=8.5 cm]{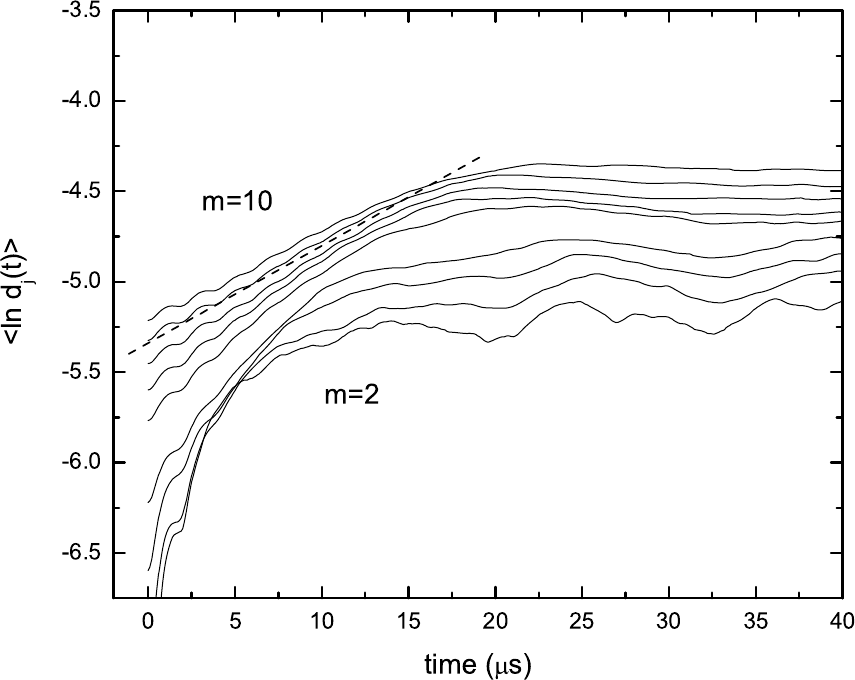}
\caption{ Average $\ln d_j(t)$ for different embedding dimensions. The best region for $\lambda_L$ has been shown by dotted line.}
\label{fig:lyap}
\end{figure}

\begin{table}
\caption{Error in the estimation of the correlation dimension and the +ve Lyapunov exponent}
\label{table:lyap}
\centering
\begin{tabular}{|c| c |c|}
\hline
\hline
data length & $D_{corr}$ &$\lambda_L$ \\
\hline
500 & 9.47 &0.240 \\
1000 & 4.00 &0.056 \\
1500 & 3.85 &0.058 \\
2000 & 3.78 &0.054 \\
2500 & 3.87 &0.058 \\
\hline
\end{tabular}
\end{table}

The presence of a +ve Lyapunov exponent ($\lambda_L$) is the most reliable signature of the chaotic dynamics and it is estimated using the Rosestein algorithm~\cite{physicaD:rosenstein}  which has been discussed in Subsection~\ref{subsection:analysis} of Section~\ref{section:experiment}. Fig.~\ref{fig:lyap} shows $<\ln d_j(t)>$ vs $i\Delta t$ for  m=2 to 10. A clear scaling region is seen at a higher m shown by $--$ line. The positive $\lambda_L$ has been identified at 0.95 and 1 mbar for some DVs and they are shown in Fig~\ref{fig:cordisch}(b) and \ref{fig:cordisch}(c) by +ve sign respectively. Figs~\ref{fig:cordisch}(b) and (c) show that $\lambda_L$ becomes positive for  283, 284, and 290 V at 0.95 mbar and for 293, 296, 300 and 305 V at 1 mbar respectively. Though $D_{corr}$ quantifies the complexity present in the system, it does not guarantee the presence of chaos, which is determined by $\lambda_L$ . At  0.89 mbar initially we have a high $D_{corr}$ [Fig~\ref{fig:cordisch}(a)], but the $\lambda_L$ is not positive in this case, implying that the system is not in chaotic state.  At higher pressures we find that in general $\lambda_L$ is +ve $D_{corr}\geq 3$, suggesting a low dimensional chaos.

$D_{corr}$ and $\lambda_L$ have been estimated for different data lengths to check the error on the estimations of the $D_{corr}$ and $\lambda_L$. The estimated $D_{corr}$ and $\lambda_L$ at different data lengths have been shown in Table~\ref{table:lyap}. Both and $\lambda_L$ and $D_{corr}$ tend to show stable results for higher data lengths (larger than 1000) as seen from Table~\ref{table:lyap}.

\begin{figure}[ht]
\center
\includegraphics[width=8.5cm]{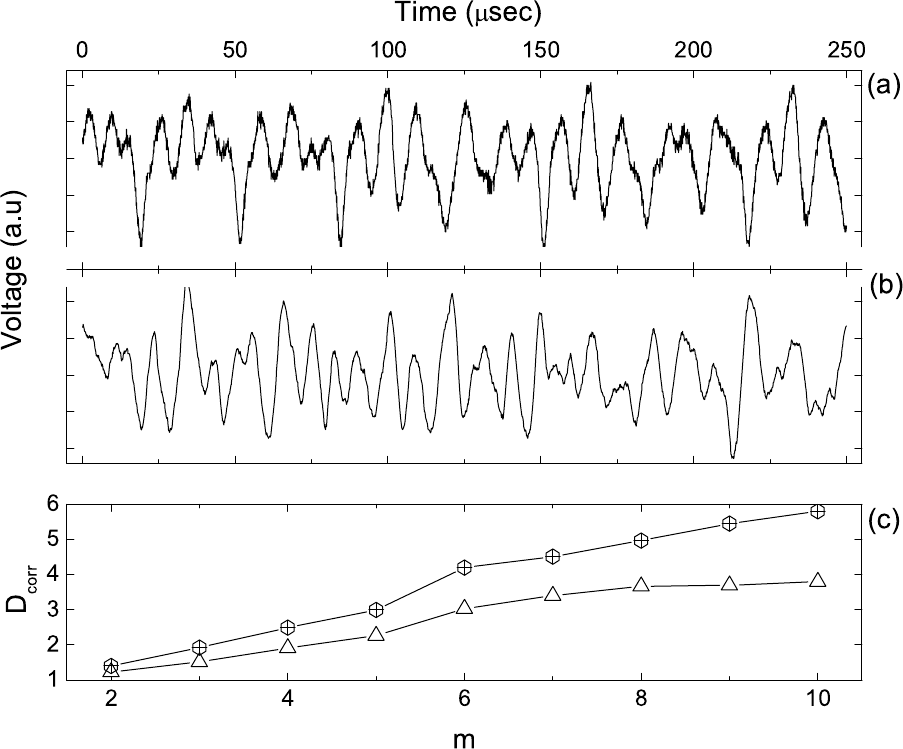}
\caption{Surrogate data using phase randomizing method: (a) original data; (b) surrogate data; (c) Correlation dimensions vs m for original ($\Delta$) and its surrogate data ($\oplus$).}
\label{fig:surgt}
\end{figure}
In order to validate the above observations, we carried out the surrogate data analysis using phase shuffled method that has been described in Subsection~\ref{subsection:analysis} of Section~\ref{section:experiment}. seen in Fig~\ref{fig:surgt}(c) The $D_{corr}$ for the original data saturates at higher m, whereas for the surrogate data  it increases with m as expected, since the surrogate data is supposed to be random in nature, and its $D_{corr}$ should be infinite~\cite{chaos:dori}. The saturation of $D_{corr}$ in our case indicates that the system dynamics is deterministic in nature.

\begin{figure*}
\centering
\includegraphics[width=13cm]{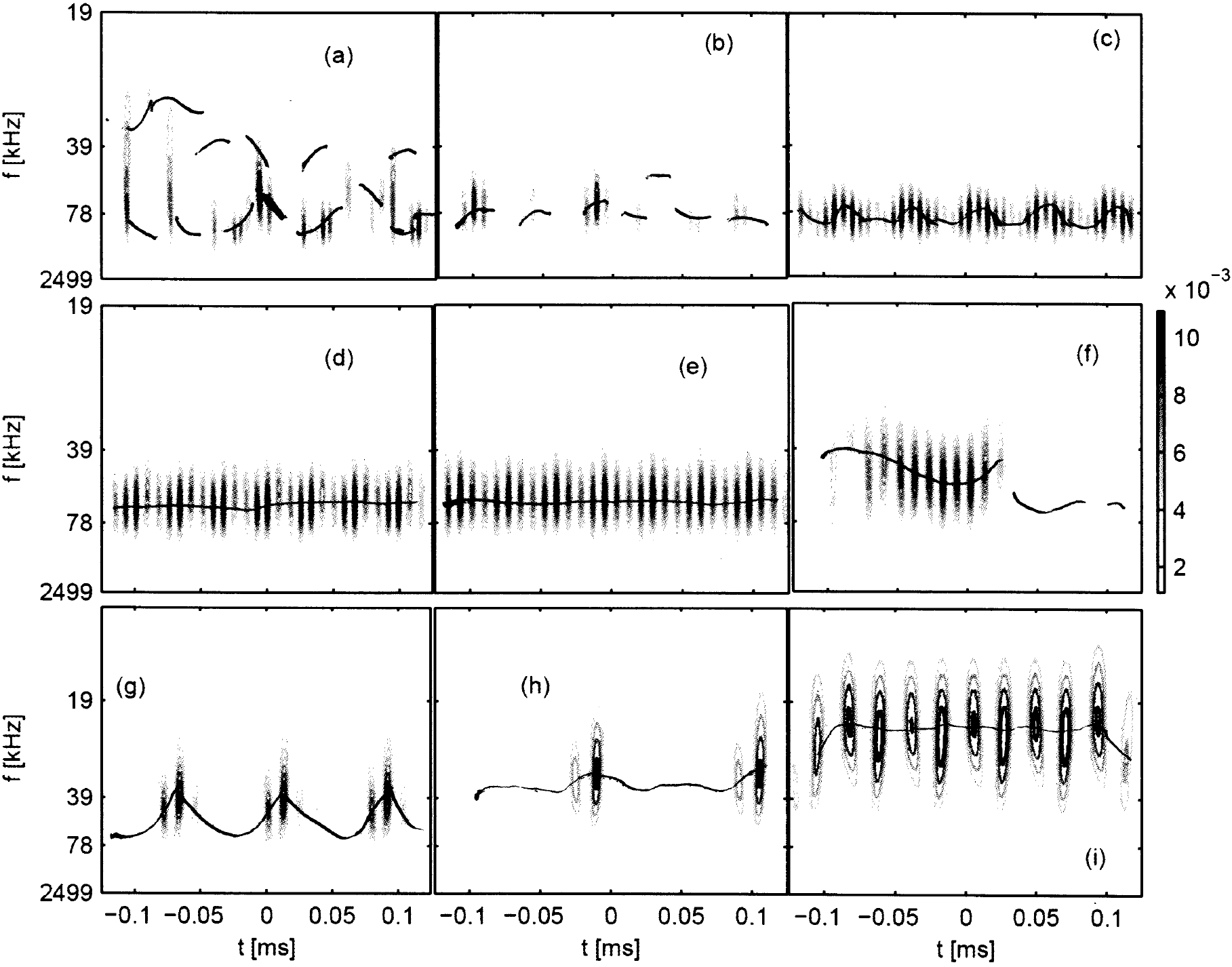}
\caption{wavelet contour plot of the fluctuations at different different DV shown in Fig~\ref{fig:raw0.95mb}. }
\label{fig:contour}
\end{figure*}
\begin{figure}
\centering
\includegraphics[width=8.5cm]{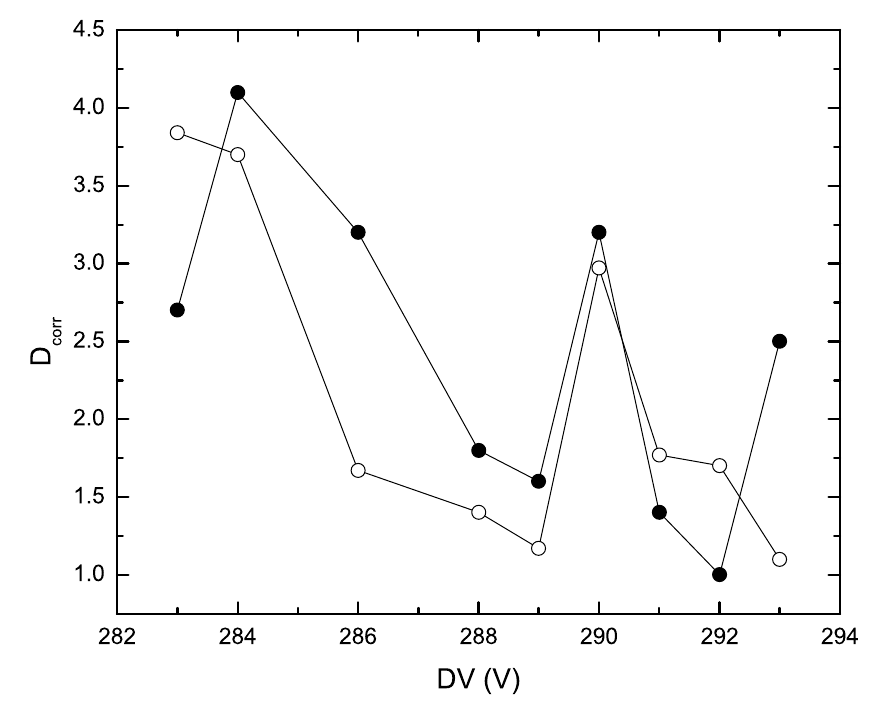}
\caption{Correlation dimension obtained from multifractal spectra ($-\bullet-$) and  from Grassberger-Procaccia algorithm (open circle).}
\label{fig:corr_mult}
\end{figure}

We also carried out a wavelet analysis of the fluctuations at 0.95 mbar and obtained the presence of chaos. The ridge plot method (Chandre, \emph{et al.}~\cite{physicaD:Chandre3}) clearly shows that  the presence of many ridges [Fig~\ref{fig:contour}(a)] indicating that at the initial stage of DV (283 V) the system is chaotic~\cite{physicaD:Chandre3}, whereas at 284 V [Fig~\ref{fig:contour}(b)] the system is weakly chaotic as only few ridges are present. With increase in DV the system becomes almost regular in nature [Fig~\ref{fig:contour}(b)$-$(e)] where ridges are approximately constant with time. But at 290 V [Fig~\ref{fig:contour}(f)] abrupt change in frequency, i.e., presence of transient features  suggest that these are the effect of nonlinearity present in the system~\cite{BAMS:Lau}. From the multifractal analysis~\cite{lecturenotes:nurujjaman} we obtained the $D_{corr}$ which is shown in Fig.~\ref{fig:corr_mult} ($-\bullet-$). This agrees quite well with the $D_{corr}$ [open circle in Fig.~\ref{fig:corr_mult}] using Grassberger-Procaccia techniques~\cite{physrevlett:grassberger,physrevA:grassberger}.

\subsubsection{Homoclinic bifurcation}
\label{subsubsection:homoclinic}
In Subsection~\ref{subsubsection:chaos}, we have reported that the fluctuations show relaxation type oscillations near the bifurcation point.

\begin{figure*}[ht]
\centering
\includegraphics{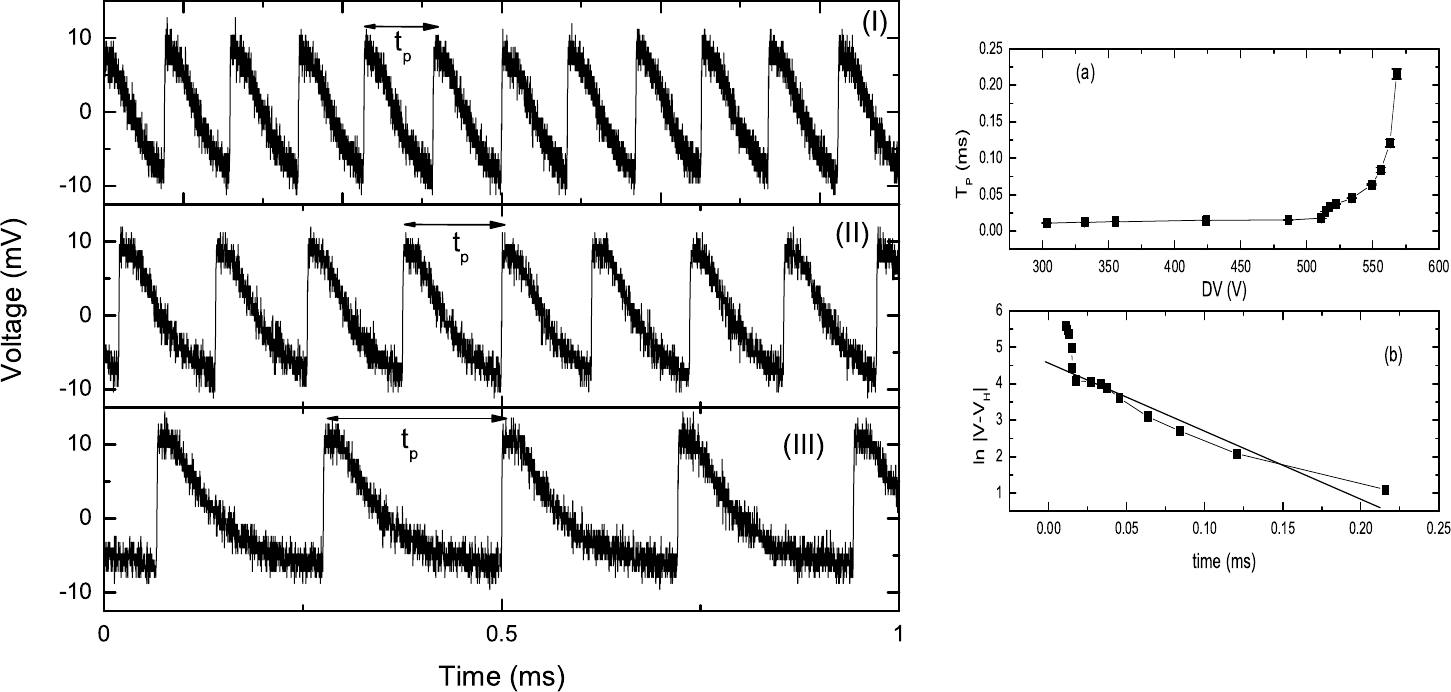}
\caption{Left panel: Timeseries showing that the period of
relaxation oscillations increases with
augmenting DV. Right panel:(a) Exponential increment of the
time period (T) with DV and (b) $\ln|V-V_H|$ vs T curve can
be fitted by a straight line indicating an underlying
homoclinic bifurcation.}
\label{fig:homoclin}
\end{figure*}

The time period (T) of these relaxation oscillations increases  with increasing DV near the bifurcation point. This eventually results in the vanishing of the limit cycle behavior beyond a critical DV ($V_H$). For larger values of DV, the autonomous dynamics exhibit a steady state fixed point behavior. Time traces from top to bottom in the left panel of Fig~\ref{fig:homoclin} depict this period lengthening of the  oscillatory behavior. A systematic  analysis of the increment in the period (T), presented in Fig.~\ref{fig:homoclin}(a) [right panel], indicates that the autonomous dynamics undergo a exponential slowing down. Consequently, the  $\ln|V-V_H|$ vs T curve can be fitted by a straight line, where $V_H$ is the bifurcation point separating the oscillatory domain and the steady state behavior. The results of Fig~\ref{fig:homoclin} indicate that the system dynamics undergo a \textbf{homoclinic bifurcation}~\cite{pre:nurujjaman} at $V_H$ resulting in the loss of oscillations. The DV ($V_H$) divides the plasma into two distinct region, The floating potential fluctuation exhibits relaxation oscillations on the one side of the $V_H$ and stable fixed point on the other side,  which is termed as being in an excitable state and is useful to study noise invoked resonance effect in glow discharge plasma that will be discussed in Section~\ref{section:resonance}.

\subsection{Conclusions}
\label{subsection:discussion and conclusion}

Glow discharges are simple systems which exhibit exotic features depending on the configuration, initial parameters, etc. Since various configurations are used in different applications like dusty plasma, plasma processing, etc. it is necessary to understand the plasma dynamics of these systems as much as possible. However the understanding of the complexities in the plasma dynamics is quite a challenging job as they arise from many degrees of freedom like different sources of free energy, different types of wave particle interaction and many other instabilities. In our present work nonlinear time series analysis has been used to quantify and differentiate complex and coherent processes at different parametric conditions. We also observed that the chaotic state driven by the DV tends to be stabilized through homoclinic bifurcation.

\section{Stochastic and coherence resonance in glow discharge plasma}
\label{section:resonance}

 \section{Introduction}
Stochastic resonance (SR) which has been observed in many physical, chemical and
biological systems ~\cite{JPhysA:benzi, arxiv:benzi,revmodphys:Gammaitoni,prl:bruce,pre:parmananda,JStatPhys:moss, prl:longtin,JPhysChem:foster,JPhysChem:Amemiya,prl:kitajo}, is a phenomenon in which the response of the nonlinear system to a weak periodic input signal is amplified or optimized by
the presence of a particular level of noise~\cite{JPhysA:benzi}, i.e.,
a previously untraceable subthreshold signal applied to a nonlinear system,
can be detected in the presence of noise. Furthermore, there
exists an optimal level of noise for which the  most efficient detection
takes place.

In 1981,  Benzi and his coworkers, had introduced the idea of the SR to
model the existence of the 100k year cycle of occurrence of the  ice age and relatively warm age
with a temperature difference of the order of 10K, which is surprisingly in phase with the Milankovitch cycle. They successfully explained the phenomena,  assuming a bistable system of the global climate,  to be separated by 10K temperature, the Milankovitch cycle as a weak periodic forcing and the short term climate fluctuations as noise, which has been discussed in Ref~\cite{JPhysA:benzi,arxiv:benzi,revmodphys:Gammaitoni}.

\begin{figure*}[ht]
\centering
\includegraphics[width=9cm]{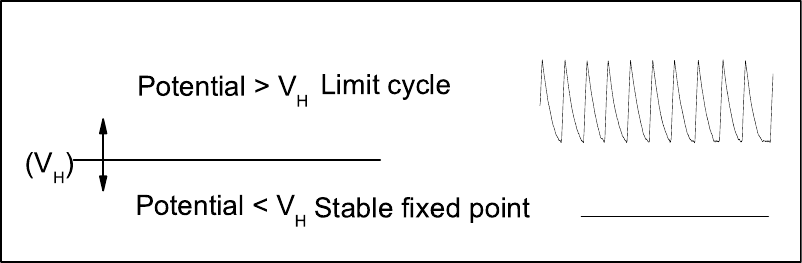}
\caption{It shows that when the control parameter crosses the potential barrier (PB), i.e., $potential >$PB system behaves
shows limit cycle behavior (up) and when ($potential<$ PB) system shows stable behavior (down).}
\label{fig:limit}
\end{figure*}

In 1993, Gang, \emph{et al.}~\cite{prl:gang}, and in 1994, Kurt Wiesenfeld, \emph{et al.}~\cite{prl:wiesenfeld}, had shown  that SR is also possible in different classes of dynamical systems based not on bistability but on excitable dynamics. They proposed a system consisting of a potential barrier and above this barrier the system shows deterministic dynamics (limit cycle oscillation) and below, stable fixed point as illustrated in Fig~\ref{fig:limit}.

The figure shows that when the control parameter (potential) crosses the PB, i.e., the potential is greater than PB, the system shows a limit cycle behavior and when the potential is less than PB, the system exhibits a stable fixed point behavior.  Now if we set the control parameter (potential) below the barrier and apply noise (stochastic perturbation on the potential) to the system and whenever the barrier is crossed, the system returns to its fixed point or ``rest state" deterministically~\cite{prl:gang,prl:wiesenfeld,pre:strogatz}, i.e., whenever the barrier is crossed, the system traverses one oscillation. So for moderate noise level we can get back relatively coherent oscillations, which is called coherence resonance (CR)~\cite{prl:pikovsky5}. Again when a subthreshold periodic signal in the form of pulse and noise are added to the system below the potential barrier the probability of crossing the barrier by noise at the time of occurrence of the periodic pulse is maximum and hence one can get back deterministic dynamics in the form of  periodic oscillations of frequency of the applied periodic pulse for optimum noise level which is the stochastic resonance of an excitable system. So for excitable systems both SR and CR are possible. Based upon excitability CR have been observed in plasma~\cite{prl:Lin I,pop:dinklage}, electronic circuits~\cite{pre:postnov}, optical systems~\cite{prl:Giacomelli,prl:avila}, and chemical systems~\cite{pre:miyakawa,pre:istavan,pre:Santos1,pre:revera} and SR has been observed in chemical systems~\cite{pre:Santos1}, human brain and many other systems~\cite{revmodphys:Gammaitoni,prl:kitajo}.

Lin I and Jeng-Mei Liu~\cite{prl:Lin I} and A. Dinklage, \emph{et al.}~\cite{pop:dinklage}, had studied CR in weakly ionized rf magnetoplasma and neon glow discharge plasma based on excitable dynamics. In both the cases, the excitability has been achieved through Hopf bifurcation for particular parametric region of the discharge. In our experiments,  we get excitable dynamics, for certain discharge parameters, in the region greater than the Paschen minimum, where the system exhibits selfsustained relaxation oscillations, and  change to excitable fixed point behavior with increase in the DV, which has been discussed in Subsection~\ref{subsubsection:homoclinic} of Section~\ref{section:chaos}. The DV at which these oscillations cease may be termed the bifurcation point ($V_H$) which act as PB in this experiment. Near The $V_H$, the time period of these relaxation oscillations increases  dramatically upon further incrementing
DV. This eventually results in the vanishing of the limit
cycle behavior beyond the critical DV ($V_H$) and exhibits excitable dynamics through homoclinic bifurcation. So below $V_H$ the system shows excitable behavior and useful to study noise invoked resonance experiments, in these regions.

\subsection{Stochastic resonance}
\label{subsection:SR}
For our experiments on stochastic resonance, the reference
voltage $V_0$ was chosen such that $V_0>V_H$ and therefore
the autonomous dynamics, by virtue of
an underlying homoclinic bifurcation, exhibit
steady state behavior. The discharge voltage $V$
was thereafter perturbed  $V=V_0+S(t)+D\xi$,
where  $S(t)$ is the subthreshold periodic pulse train
chosen for which $V=V_0+S(t)>V_H$, (subthreshold
signal does not  cause the  system to cross over
to the oscillatory regime) and $D\xi$
is  the added Gaussian white noise $\xi$ with
amplitude  $D$. Subthreshold periodic square pulse of
width $20~\mu s$ and duration 2 ms was constructed
using Fluke PM5138A function generator. Meanwhile,  the
gaussian noise  produced using the HP 33120A noise
generator was subsequently amplified using a noise
amplifier.

\begin{figure*}[ht]
\centering
\includegraphics[width=5.5in]{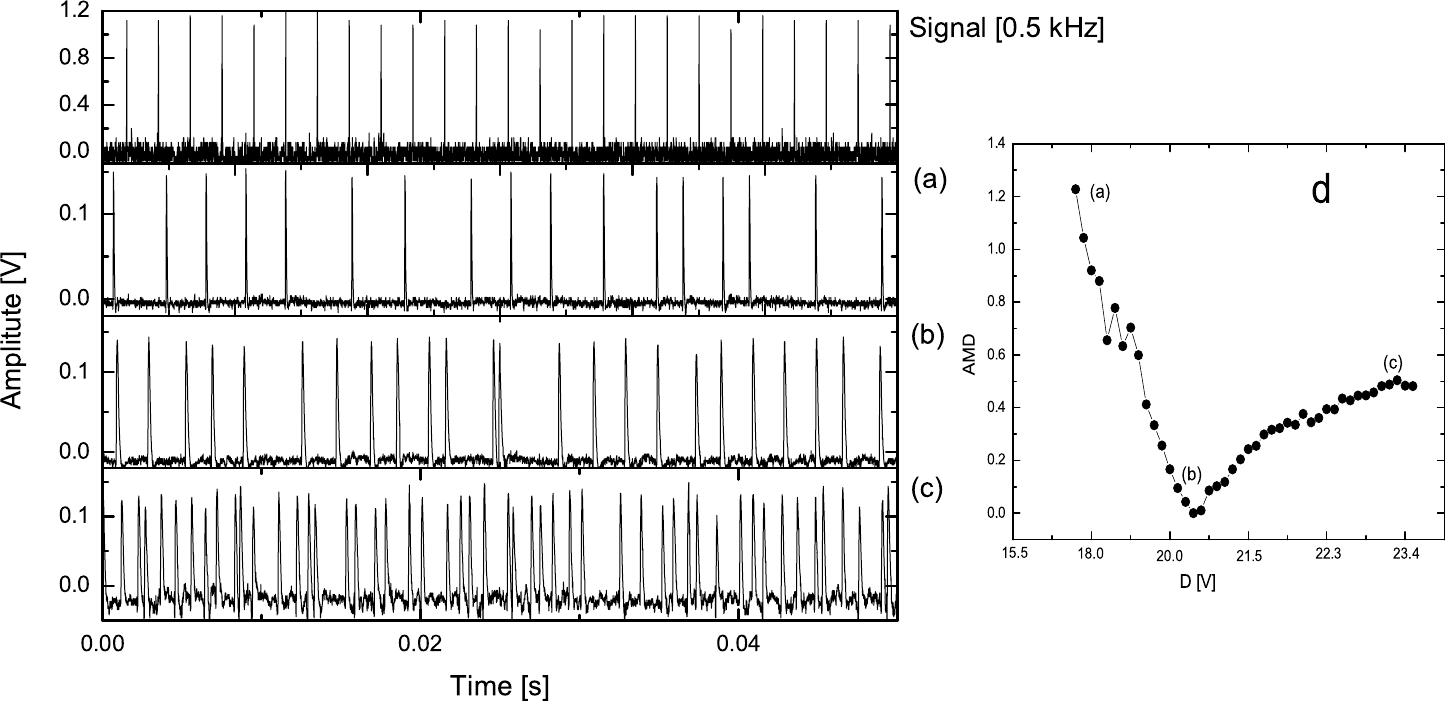}
\caption{Noise induced system response in the floating potential
fluctuations for low, medium, and high amplitude noise in
conjunction with a subthreshold periodic square pulse. The right
panel [Fig~\ref{fig:periodic}] shows the AMD as a function of
noise amplitude for the experiment performed at $V_0=307$ V and
pressure= 0.39 mbar. Left panel shows the subthreshold periodic
pulse train and the  three time series of floating
potential fluctuations at low level noise (a); at optimum
noise value (b) and at high amplitude noise (c).}
\label{fig:periodic}
\end{figure*}

Fig.~\ref{fig:periodic}$(a)-(c)$ show time series of the system response  in the presence of an identical subthreshold signal for three different amplitudes of imposed noise. The subthreshold periodic pulse train is also plotted, in the top most graph of the left panel, for comparison purposes. Fig.~\ref{fig:periodic}(a) shows that there is little correspondence between the subthreshold
signal and the system response for a low noise amplitude.
However, there is excellent correspondence at an intermediate noise
amplitude [Fig.~\ref{fig:periodic}(b)]. Finally,  at higher
amplitudes of noise the subthreshold signal is lost amidst
stochastic fluctuations of the system
response [Fig.~\ref{fig:periodic}(c)]. Absolute
mean difference (AMD) which has been discussed in Subsection~\ref{subsection:analysis} of Section~\ref{section:experiment},  used to quantify the
information transfer between the subthreshold
signal and the system response, is defined
as $AMD=abs(mean(\frac{t_p}{\delta}-1))$.
$t_p$ and $\delta$ are the inter-peak
interval of the response signal
and mean peak interval of the subthreshold periodic
signal respectively.  Fig~\ref{fig:periodic}(d) shows
that the experimentally computed AMD versus noise amplitude D curve
has a unimodal structure typical for the SR phenomena.
The minima in this curve  corresponds to
the optimal noise level for which maximum information transfer
between the input and the output takes place.
\subsection{Coherence resonance}
\label{subsection:CR}

\begin{figure*}[ht]
\centering
\includegraphics [width=5.5in]{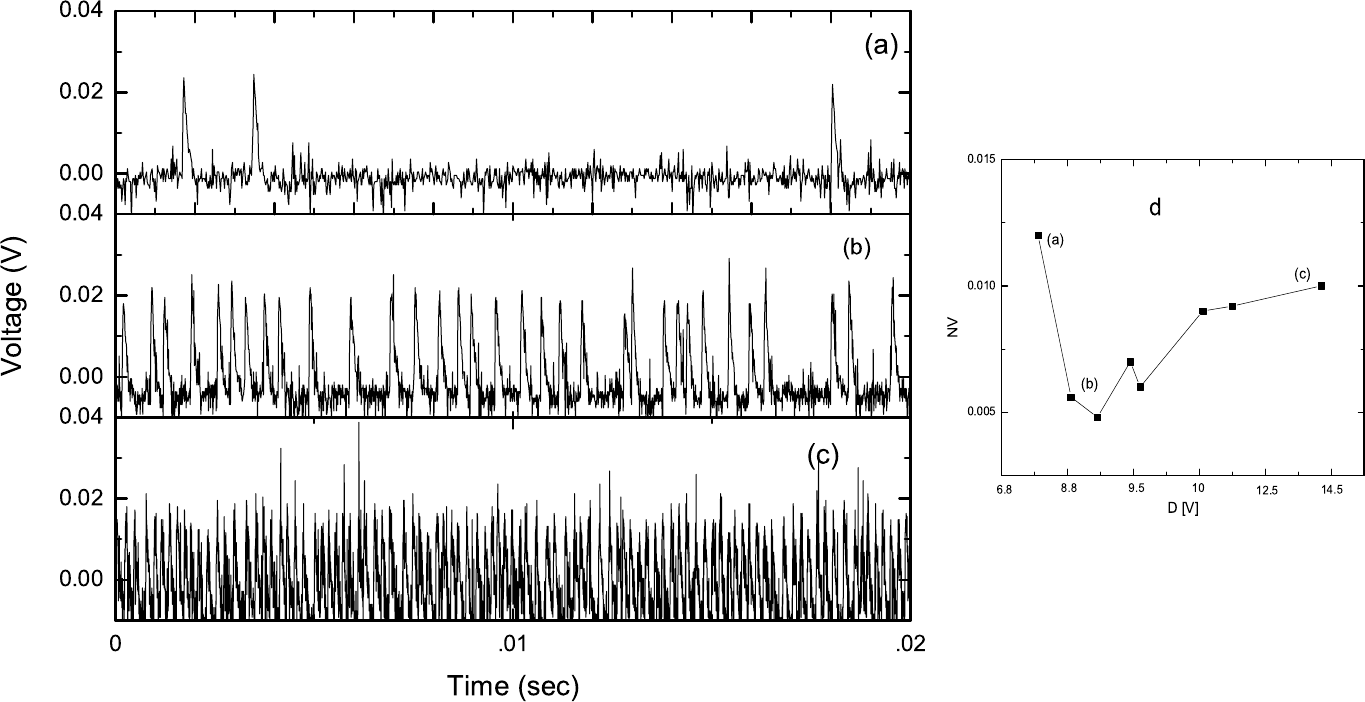}
\caption{Emergence of coherence resonance: The right
panel  shows the NV as a function
of noise amplitude for the experiments performed
at $V_0=344$ V and pressure = 0.5 mbar. Left panel: The
time series of  the floating potential fluctuations
for (a) low noise, (b) optimum noise and (c) high level noise}
\label{fig:NV}
\end{figure*}

For the experiments on coherence resonance
DV $(V_0$) was located such that
the floating potential fluctuations
exhibit  fixed point behavior.
In order to minimize the effect of parameter drift,
a set-point ($V_0$) quite far from the homoclinic
bifurcation ($V_H$) was chosen. Subsequently, superimposed
noise on the discharge voltage was increased and the
provoked dynamics analyzed. The normalized variance (NV) which has been discussed in subsection~\ref{subsection:analysis} of Section~\ref{Section:experiment} was
used to quantify the extent of induced
regularity. It is defined as $NV=std(t_p) / mean(t_p)$,
where $t_p$ is the time elapsed between successive peaks. It
is evident that  more regular the induced
dynamics the lower the value of the computed NV.
For purely periodic dynamics the NV goes to zero.

Fig.~\ref{fig:NV}$(a)-(c)$ (left panel)
show the time series of the floating potential fluctuations
for different noise levels and Fig~\ref{fig:NV}(d) (right panel)
is the experimental NV curve as a function of noise
amplitude D. The point (a) in Fig~\ref{fig:NV}(d)
(time series shown in Fig.~\ref{fig:NV}(a)) is associated
with a low level of noise where the activation
threshold is seldom crossed, generating a sparsely
populated irregular spike sequence. As the noise
amplitude is increased, the NV decreases,
reaching a minimum (b) in Fig~\ref{fig:NV}(d)
(time series shown in Fig.~\ref{fig:NV}(b))
corresponding to an optimum noise level where maximum regularity of
the  generated spike sequence is observed.  As the
amplitude of superimposed noise is increased
further, the observed regularity is destroyed
manifested by an increase in the
NV; label (c) in Fig~\ref{fig:NV}(d)
(time series shown in Fig.~\ref{fig:NV}(c)).
This is a consequence of the dynamics being dominated by noise.

\subsection{Discussion}

\begin{figure}[ht]
\centering
\includegraphics [width=3.5in]{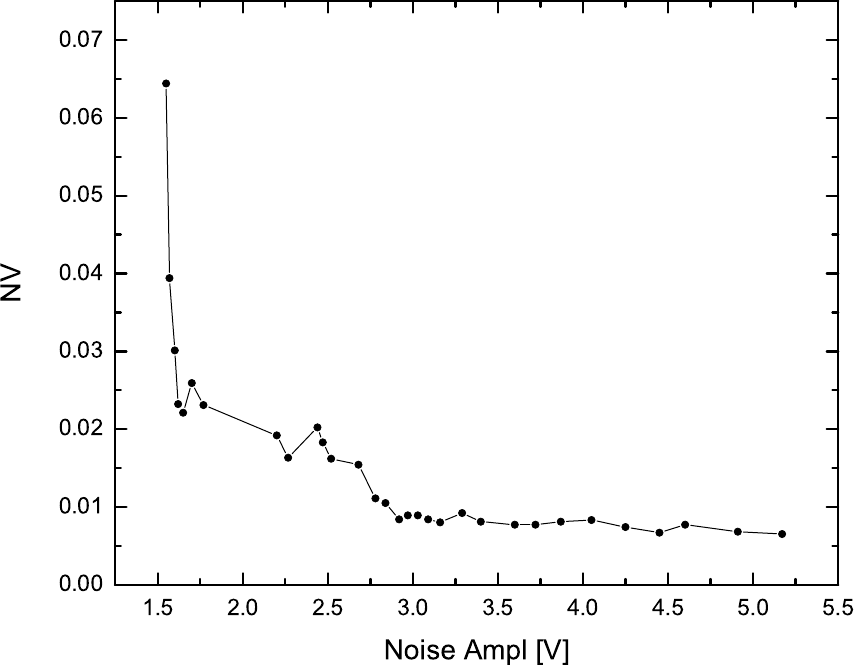}
\caption{Emergence of coherence resonance:  NV as a function
of noise amplitude for the experiments performed.
at $V_0=344$ V and pressure = 0.5 mbar.}
\label{fig:NVexp}
\end{figure}

\begin{figure*}[ht]
\center
\includegraphics[width=11cm]{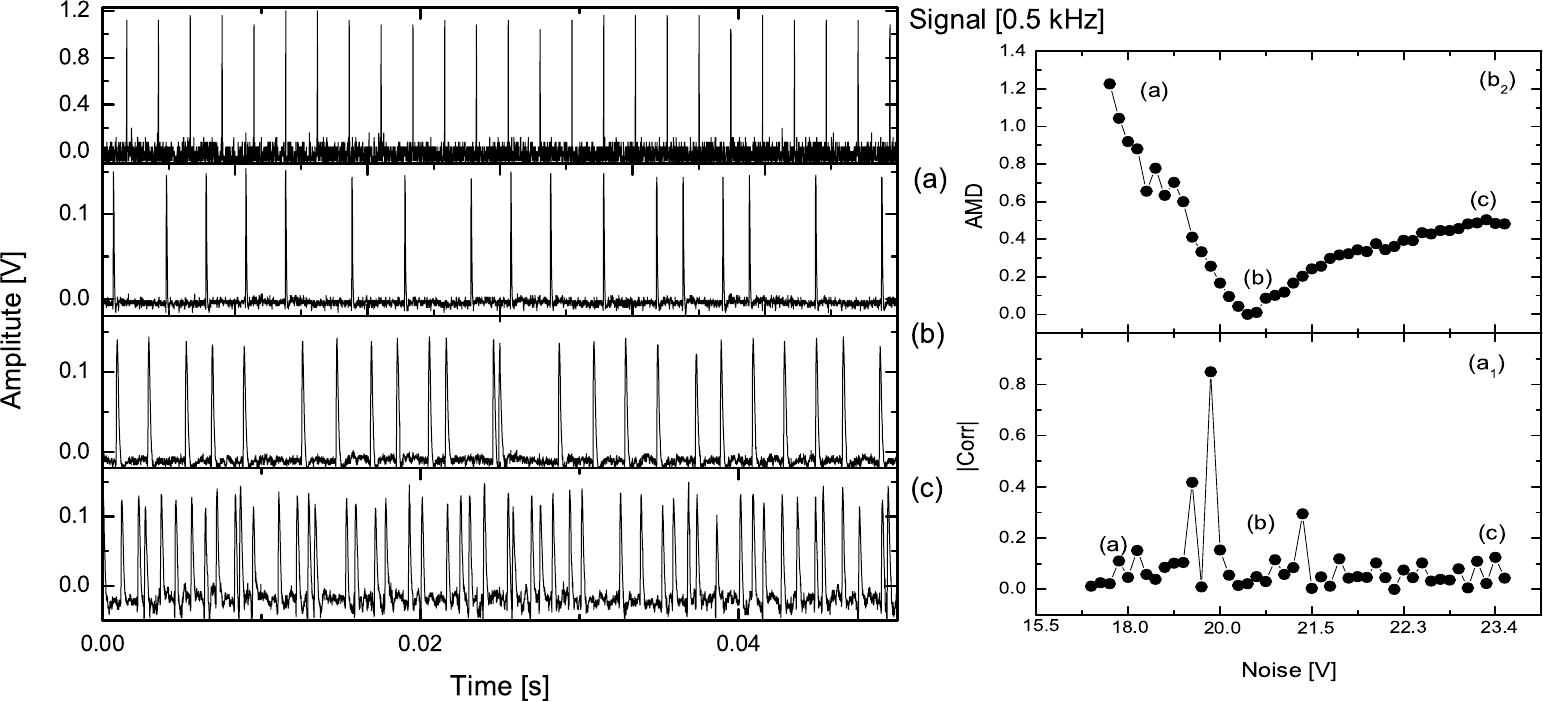}
\caption{$a_1$: $C_0(0)$ vs applied noise amplitude. $b_1$ is the corresponding AMD. Fig$a_1$ does not show any maximum at optimum noise (b)
which is clear from $b_1$.}
\label{fig:crAMD}
\end{figure*}

The effect of noise has been studied experimentally near a homoclinic bifurcation in glow discharge plasma system~\cite{pre:nurujjaman}. Our study demonstrates the  emergence of SR for
periodic subthreshold square pulse signals and the induction of CR via purely stochastic fluctuations.
In SR experiments, the efficiency of information transfer was quantified using AMD instead of the  power
norm which has been utilized elsewhere ~\cite{pre:parmananda}. The advantage of using  this method in comparison to  the power norm ($C_0(0)$)~\cite{pre:parmananda} lies in the fact that AMD remains independent of the lag between the measured floating potential and the applied periodic square pulse. This is of relevance to our experimental system, where  invariably there exists  a lag, at times
varying in time due to the parameter drifts. Comparison between the estimated $C_0(0)$ and AMD
have been shown in Fig~\ref{fig:crAMD}. It is obvious from Fig~\ref{fig:crAMD}($a_1$) that $C_0(0)$ does not show any peak at optimum noise level. Whereas, AMD shows nice agreement with regular spiking of the signal [Fig~\ref{fig:crAMD}($b_1$)]. For the CR experiments it was occasionally observed that while with an  initial increase in noise amplitude (D) NV reaches  a minimum, the subsequent rise of
NV for even higher amplitudes of noise was suppressed as shown in Fig~\ref{fig:NVexp}.
This leads to the modification of the unimodal profile, a signature of the CR phenomenon.

A possible explanation for this suppression is that by virtue of the superimposed high frequency
noise (bandwidth 500 KHz) and fast responding internal plasma dynamics, the system has the
capability of exciting high frequency regular modes within  the ion plasma frequency (105 kHz). This in
turn  leads to the persistence of low NV values. Finally, in Refs~\cite{prl:Lin I,pop:dinklage} both
the destructive and constructive role of noise (CR only) have  been reported  for glow discharge and magnetized rf discharge plasma systems respectively. However, both these experiments were carried out in the vicinity of the Hopf bifurcation.  In contrast, for the present work  we studied both SR and CR in the neighborhood of the homoclinic bifurcation.

\section{Realization of SOC behavior in glow  discharge plasma}
\label{section:soc}

\section{Introduction}

Many empirical analyses suggest that power-law behavior in the distribution of some quantities are quite frequent in nature  for example, $1/f$ noise, Gutenberg-Richter law for the magnitudes of earth quakes, stock market, forest fires, ecology, weather etc~\cite{geophys:pbak,prl:carlson,sciAm:pbak,Physletta:pbak,prl:dhar}. From the point of equilibrium statistical mechanics, power law is considered as a signature of critical fluctuations observed near a second order phase transition. As we approach the critical temperature, correlation length diverges and measurable quantities exhibit power law. However, power laws are often seen in many other systems, not necessarily connected with equilibrium situations, some of these systems are driven systems which are very different from equilibrium systems. In the case of second order transition, the critical state is reached by only through parameter tuning. The idea of self organized criticality (SOC) introduced by  Bak, Tang and Wiesenfeld~\cite{PRL:PBak} was that when the systems are driven away from equilibrium, they reach a state near to, but not at, the state that is marginal to major disruptions without any significant tuning of the system from outside. These systems are characterized by a spectrum of spatial and temporal scales of the disruption events that exists in remarkably similar form in a wide variety of different physical systems.

The simple concept of SOC is the sandpile. Consider, the process of building a sandpile by dropping sand on a particular spot. As we add sand, the pile will grow and always chooses its shape which is independent of the builder. This property is called self organization. The sandpile surface will steepen until its slope reaches the angle of repose, the critical angle beyond which further addition of sand will cause cascades of sand to topple down the sandpile. Most parts of the sand pile have a slope very close to this critical angle. Any further addition of the sand grains will trigger cascades of sand particles which are more commonly known as avalanches. The sand pile is  a good example of a critical state, where the correlation length of perturbations extends over the entire sandpile, i.e., independent of the size of the pile, an addition of one grain of sand close to the top of the pile can influence grains at the bottom. In this process there is no external parameter needs to be tuned because the critical state in the sand pile is attained as a consequence of the slow addition of the sand grains and their spatial redistribution by avalanches~\cite{PRL:PBak,book:bharati}.

The  SOC concept has been quite rigorously deployed to explain some of the turbulent transport observations in magnetically confined fusion devices like Tokamaks.  If we consider free energy source in the plasma is the gradient of the density. When this gradient becomes greater than some critical value depending upon plasma parameters, by fueling the plasma and turbulence turbulence develops around the location. This eddy flatten the density of that position and free energy moves to its adjacent place and in the similar way eddies may be triggered the new location. In this way the density profile is modified again and the process continues and produces avalanche-like phenomenon which has been successfully explained by SOC concept.

Though SOC concept has been applied quite successfully in high temperature plasmas like tokamak~\cite{PRL:BACarreras00,POP:BACarreras0,PRL:FSattin}, we first observed that the low temperature glow discharge plasmas also exhibits SOC behavior, for the region less than the Paschen minimum which has been described in Subsection~\ref{subsection:experiment} of Section~\ref{section:experiment}.

From the the Paschen curve [Fig~\ref{fig:paschen} of Section~\ref{section:experiment}] for our system we have seen that, at low pressure i.e., for less the than Paschen minimum, discharge struck at higher discharge voltage (DV) [subsection~\ref{subsection:experiment} of Section~\ref{section:experiment}]. In the lower region,  keeping the DV at a constant value of 800 V, a discharge was struck with a very faint glow at $\sim9\times10^{-3}$ Torr, and then as the filling pressure was gradually increased by means of the needle valve, the intensity of the glow suddenly became bright at about $1.6\times10^{-2}$ Torr.
\begin{figure}[ht]
\centering
\includegraphics [width=8.5cm] {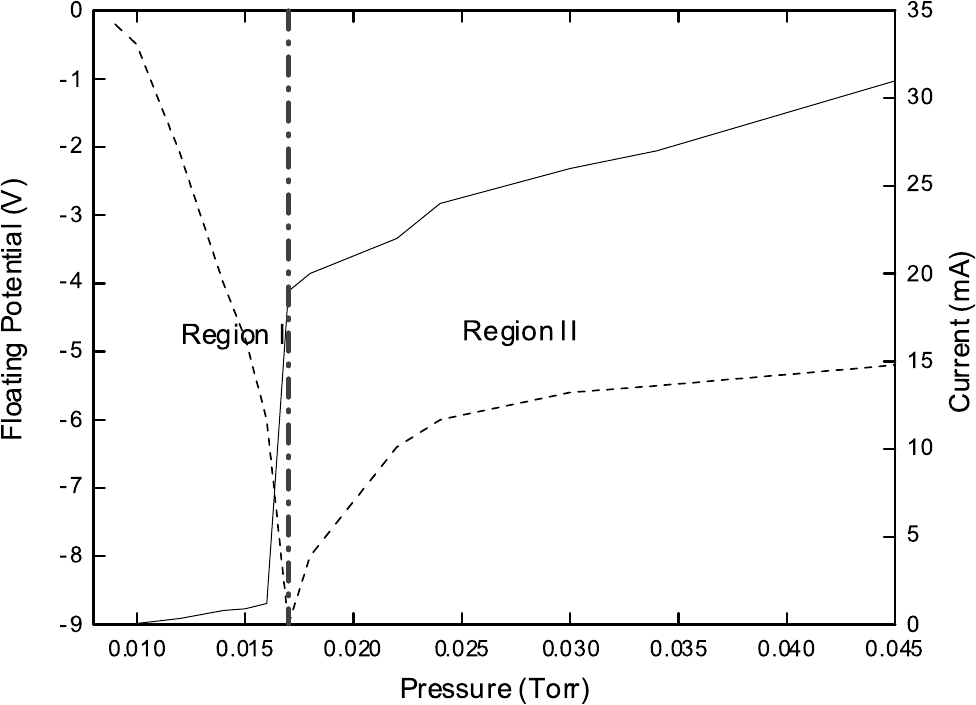}
\caption{The variation of plasma discharge current (solid line), and plasma floating potential (dotted line) with pressure. }
\label{fig:2}
\end{figure}
\begin{figure}[h]
\centering
\includegraphics [width=8.5cm] {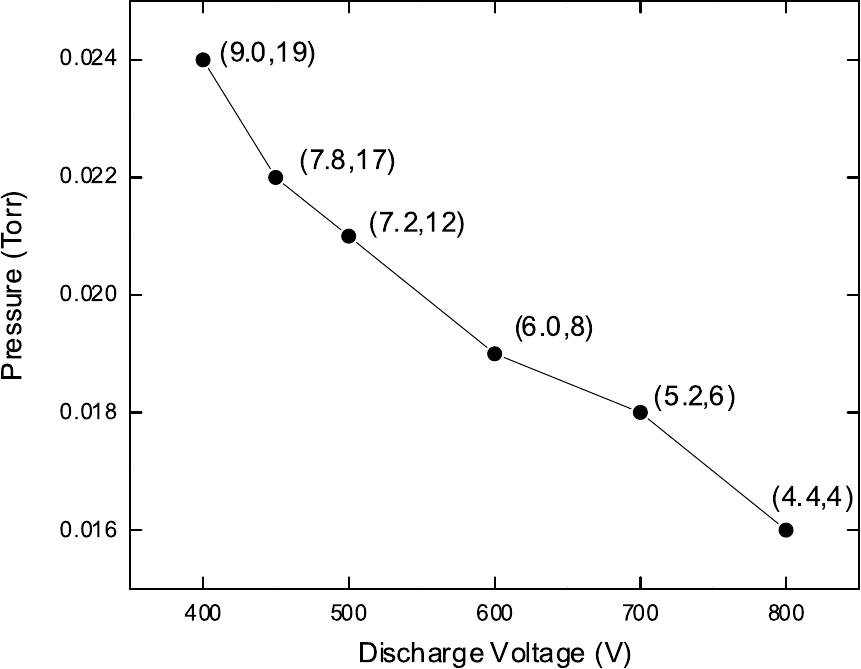}
\caption{Plot of critical pressures with discharge voltages. Corresponding floating potential, and discharge are within brackets. The first value in the brackets is the floating potential, and the second one is the discharge current in (V, mA) unit.}
\label{fig:3}
\end{figure}
 The sudden change in the plasma dynamics at the critical pressure led to a change in the floating potential, and the discharge current.  The variation of  the plasma floating potential, and the plasma discharge current with pressure have been shown in Figure~\ref{fig:2}. It shows that, the current (solid line) jumped to a larger value by a factor of 15 at $1.6 \times 10^{-2}$ Torr, and then increased gradually with pressure.  On the other hand, the floating potential rapidly fell to a negative value ($\approx -9$ V) up to the same critical pressure ($1.6 \times 10^{-2}$ Torr), and then again increased (dotted line in Fig.~\ref{fig:2}) with further increase in pressure, until it finally settled down to $\approx -5$  V. So the critical pressure divides the scanned region into two regions I and II, shown by a vertical line ($-.-$)  in Figure~\ref{fig:2}.  The plasma density, temperature, and electron-electron collision mean free path in region II are $10^7-10^8cm^{-3}$, $2-4$ eV, and $2.6\times10^5$ cm respectively. However, in  region I, it was almost impossible to obtain the I-V characteristics because of their extremely low values. Qualitatively, looking at discharge current, and glow intensity, region II is probably a normal glow discharge region, while region I might be the dark or subnormal glow discharge region. In the region I system exhibits SOC behavior.  The boundary between the two region (critical pressure) where the transition takes place is not a fixed point, but decreases with the discharge voltage as seen in  Figure~\ref{fig:3}. It is also seen that the floating potential, and the discharge current also decrease with discharge voltage.
 \begin{figure}[ht]
\centering
\includegraphics [width=8.5cm] {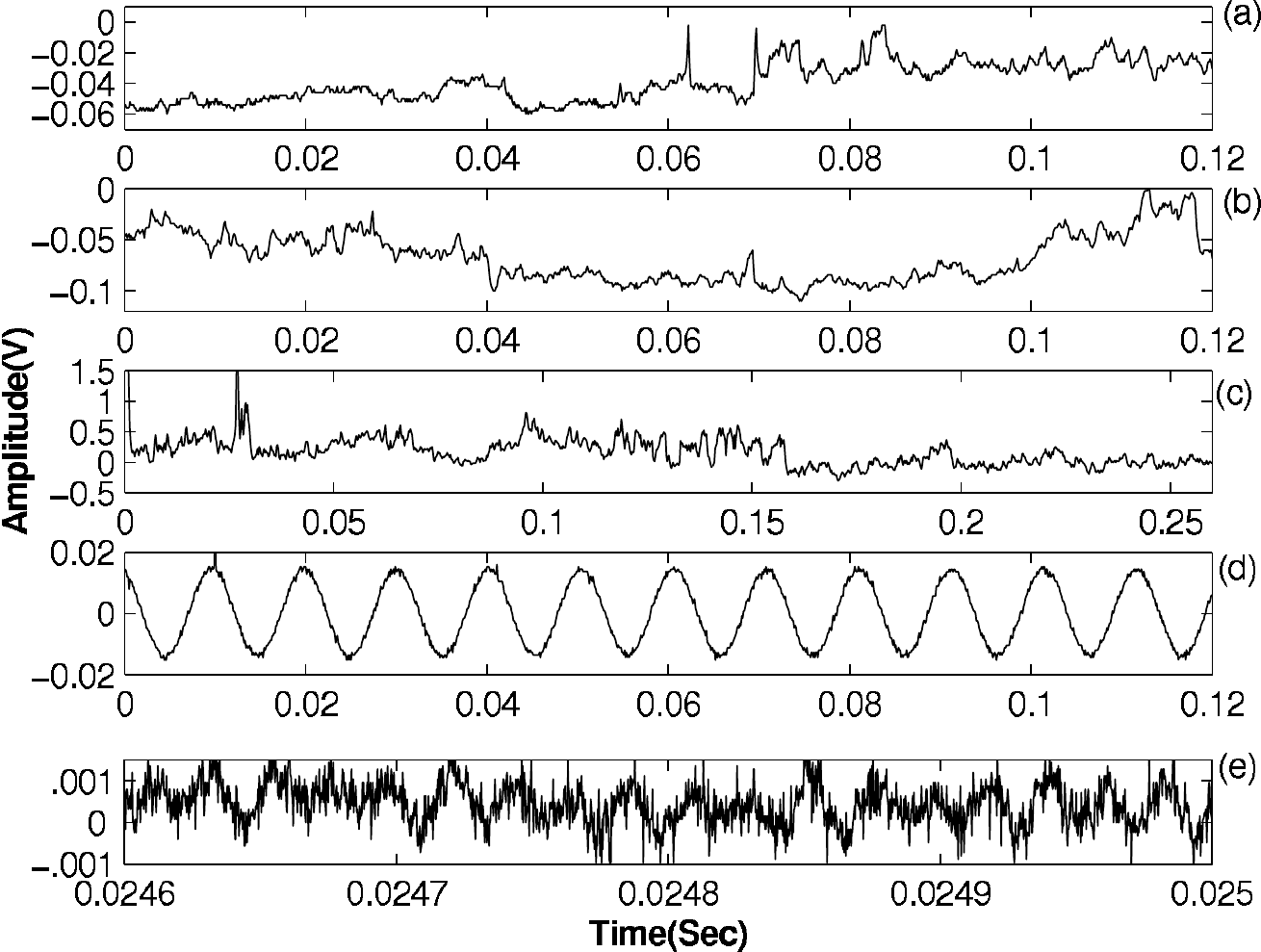}
\caption{Figure shows the electrostatic fluctuations at pressures $0.9\times10^{-2}$ (a), $1.2\times10^{-2}$ (b),  $1.5\times10^{-2}$ (c), $1.7 \times10^{-2}$  (d), and $2.2\times10^{-2}$ Torr (e) respectively.}
\label{fig:4}
\end{figure}

Figures~\ref{fig:4}(a), (b), and (c) are the typical electrostatic fluctuations at pressures $0.9\times10^{-2}$, $1.2\times10^{-2}$, and $1.5\times10^{-2}$ Torr respectively in region I, while (d), and (e) are the fluctuations at pressures $1.7 \times10^{-2}$ and $2.2\times10^{-2}$ Torr respectively in region II. In the next section it will be shown using different analysis techniques that the fluctuation in the region I is consistence with SOC behavior.

\subsection{Analysis of SOC behavior}
\label{subsec:SOC}

The experimental evidences considered as main ingredients  of SOC are $1/f^{\beta}$ ($\beta>$0) power law (where $f$ is the frequency of the fluctuations obtained from fast Fourier transform) ~\cite{PRL:PBak,PhLettA:TLRhodes,PhyslettA:Skokov,PRL:Kim}, long-range correlation~\cite{POP:BACarreras2}, and nongaussian probability distribution function (PDF)~\cite{conf:Xu}. From the power spectral analysis which have been discussed in Subsection~\ref{subsection:analysis} of Section~\ref{section:experiment}, we have estimated the $\beta$ from ln(Power) versus ln($f$).

 For long-range time correlation we estimated the Hurst exponent H using rescale range method, and the exponent ($\alpha$) of autocorrelation function (ACF) decay, as described below. Both the techniques have been discussed in Subsection~\ref{subsection:analysis} of Section~\ref{section:experiment}.

 \begin{figure}[ht]
\centering
\includegraphics [width=8.5cm]{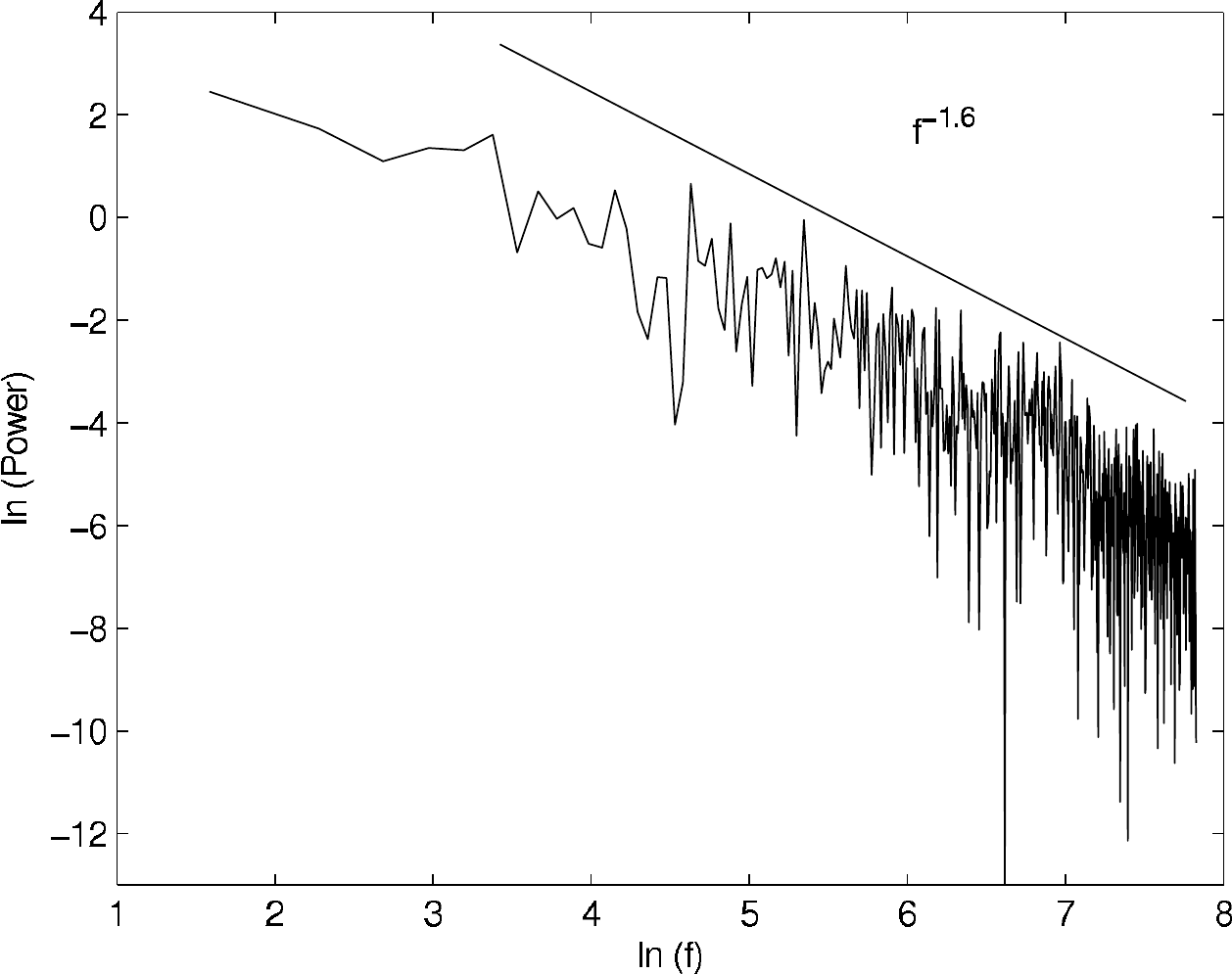}
\caption{ln(power) vs. ln $f$ plot. The solid line shows the best fit.}
\label{fig:5}
\end{figure}

 \begin{figure}[ht]
\centering
\includegraphics [width=8.5cm] {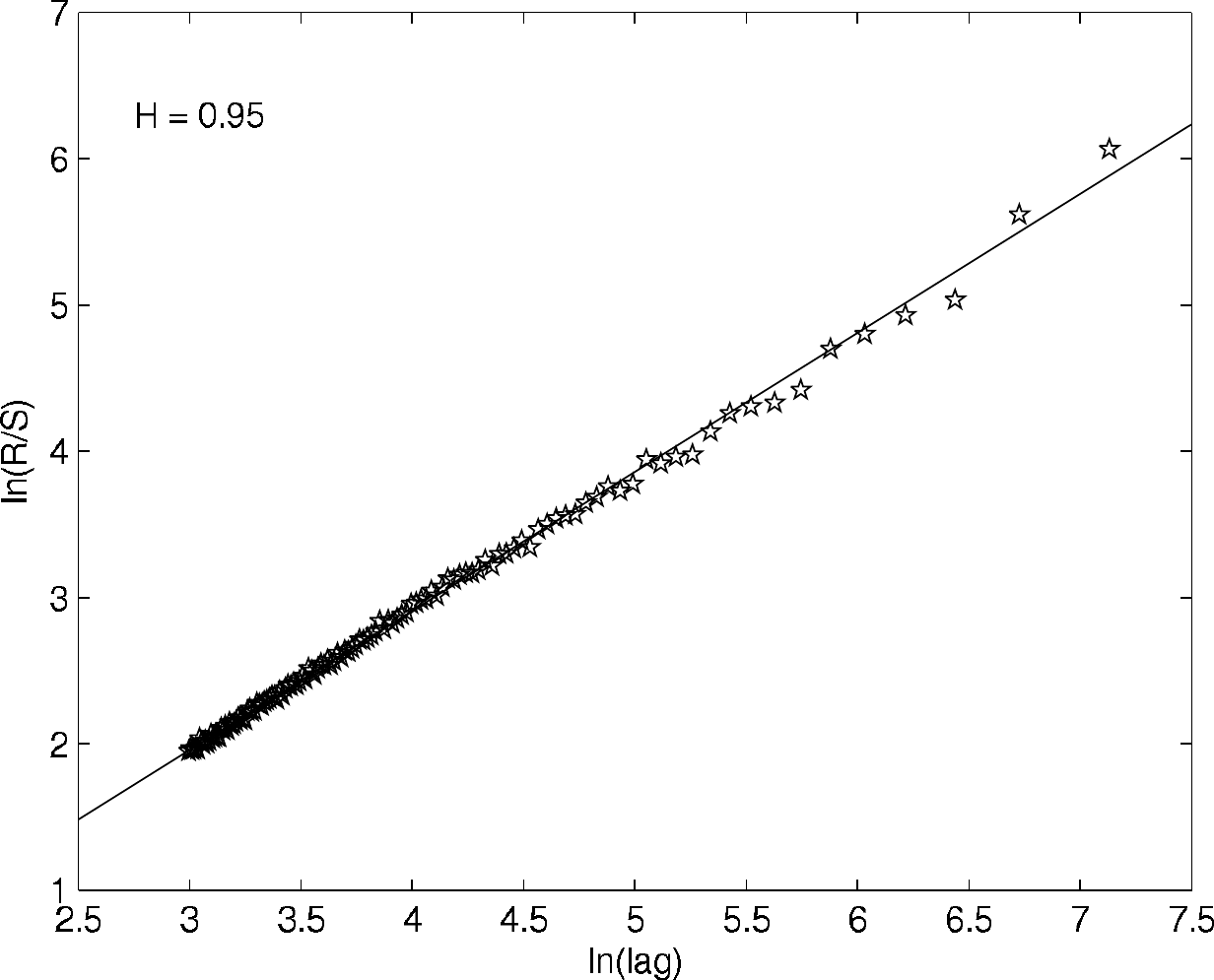}
\caption{ln(R/S) vs. ln(lag) plot for the electrostatic fluctuation at the pressure $1.4\times10^{-2}$ Torr. The solid line shows best fit.}
\label{fig:6}
\end{figure}

\begin{figure}[ht]
\centering
\includegraphics [width=8.5cm] {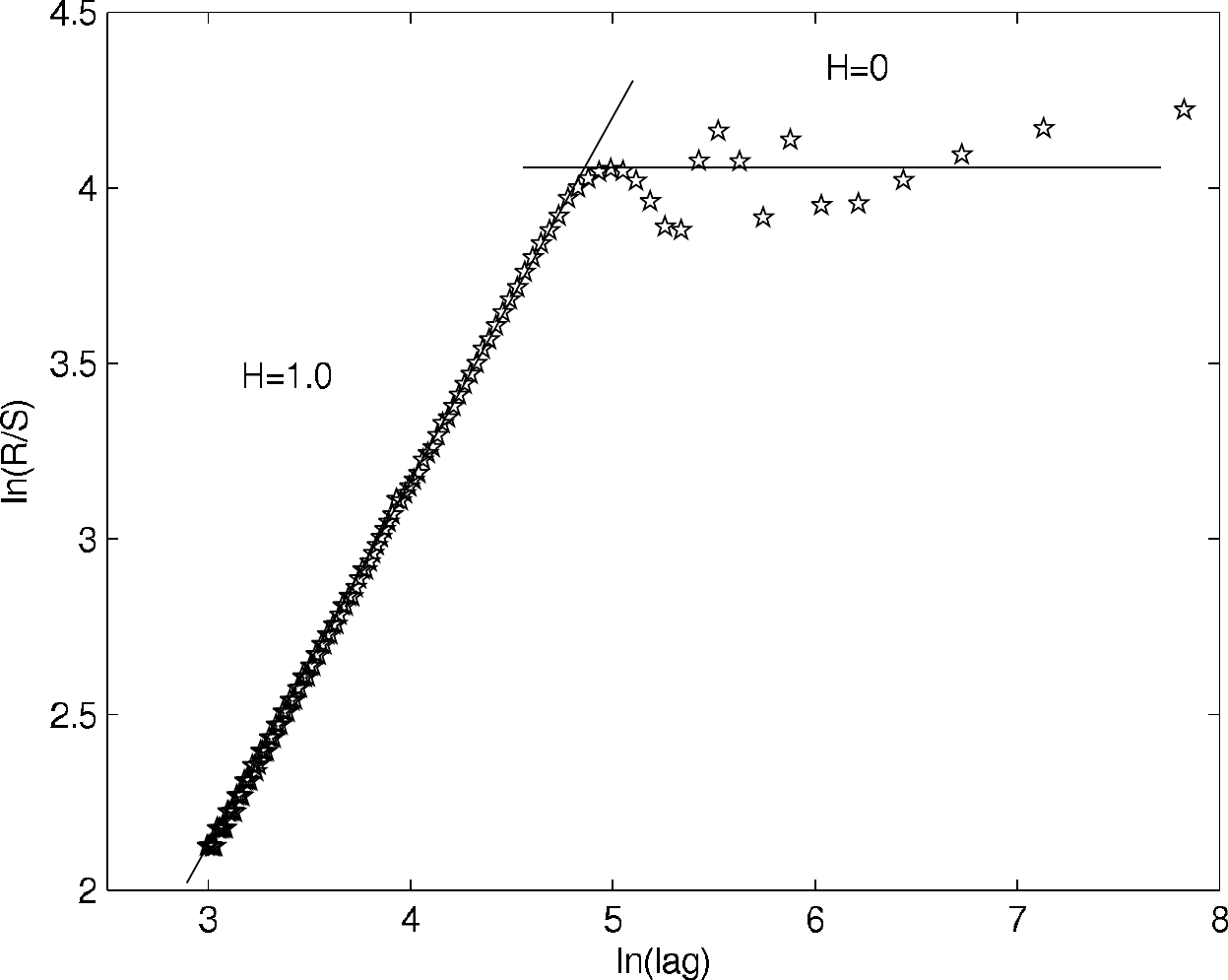}
\caption{R/S as a function of time lag  for the electrostatic fluctuation of the coherent oscillations at the pressure $1.7\times10^{-2}$ Torr. For one time period lag, H=1, and for lag more than one period, H=0.}
\label{fig:7}
\end{figure}

\begin{figure}[ht]
\centering
\includegraphics [width=8.5cm] {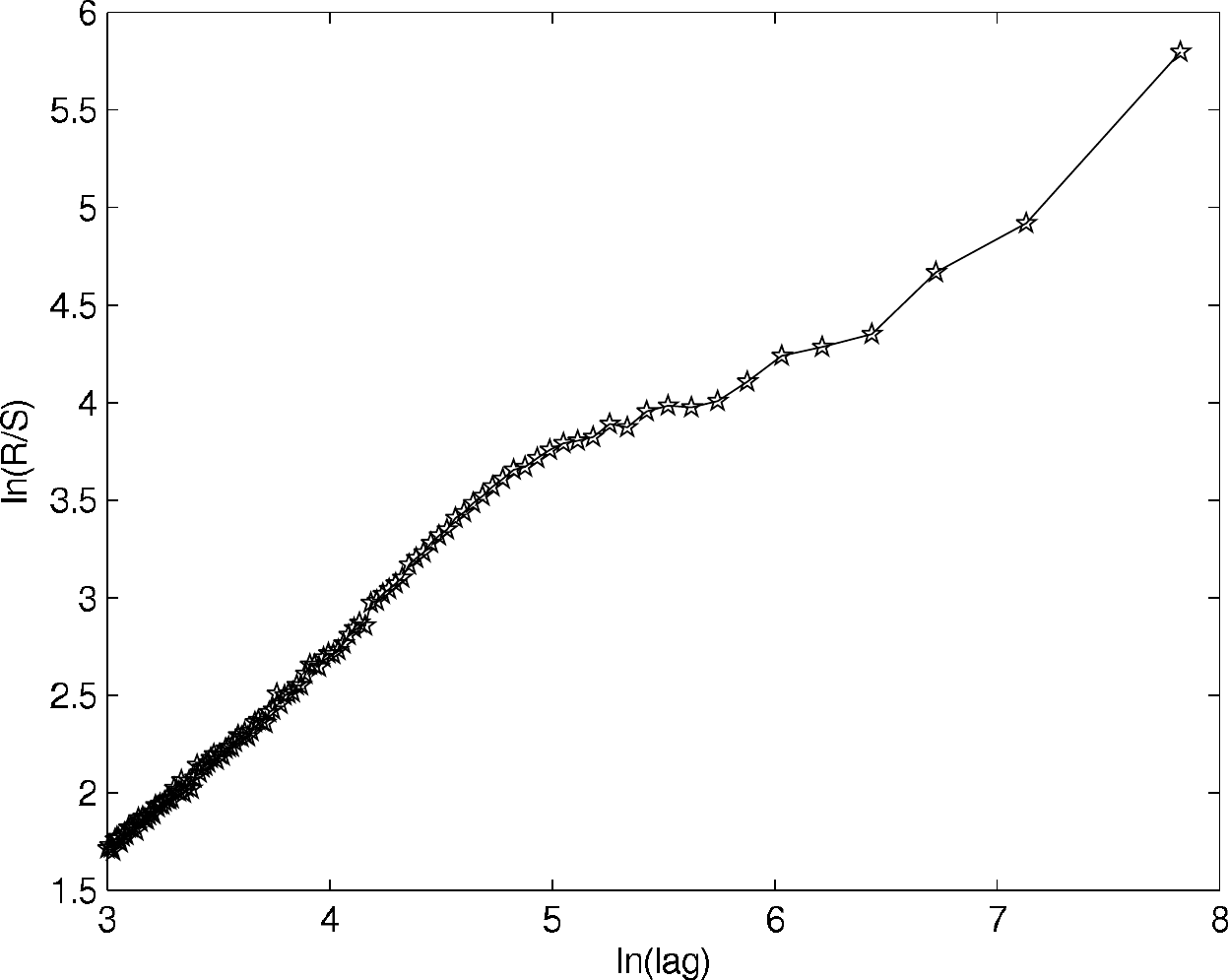}
\caption{ln(R/S) versus ln(lag) plot at the pressure $2.2\times10^{-2}$ Torr, more than one slope indicates instabilities with many modes.}
\label{fig:8}
\end{figure}

 In order to verify nongaussianity we obtained the PDF of the fluctuating data.

\begin{table}[h|b|p|]
\center
\caption{{\small In the following table the decay exponent $\alpha$ of the ACF, H from ACF, H using R/S, and the power spectral index $\beta$ have been shown in the second, third, fourth, and fifth column respectively, for the pressures shown in the first column.} }
\begin{tabular}{c c c c c}
\hline
pressure & $\alpha$  & Hurst   & Hurst  & $\beta$  \\
$ 10^{-2}$(Torr) & (ACF)& (ACF)  & (R/S) & (PS)\\
\hline
\hline
0.9 & 0.38 & 0.81 & 0.97 & 1.85\\
1.2 & 0.34 & 0.83 & 0.96 & 1.65\\
1.5 & 0.23 & 0.88 & 0.95 & 1.60\\
\hline
\end{tabular}
\label{table1}
\end{table}

Figure~\ref{fig:5} shows ln (Power) vs. ln $f$ of the fluctuations in region I, from which we estimated the exponent to be $\approx 1.6$.  This  agrees well with the numerical ~\cite{PRL:PBak,PRA:PBak}, as well as experimental observations ~\cite{PhLettA:TLRhodes,PhyslettA:Skokov,POP:BACarreras2} in the presence of SOC behavior.

\begin{figure}[ht]
\centering
\includegraphics [width=8.5cm] {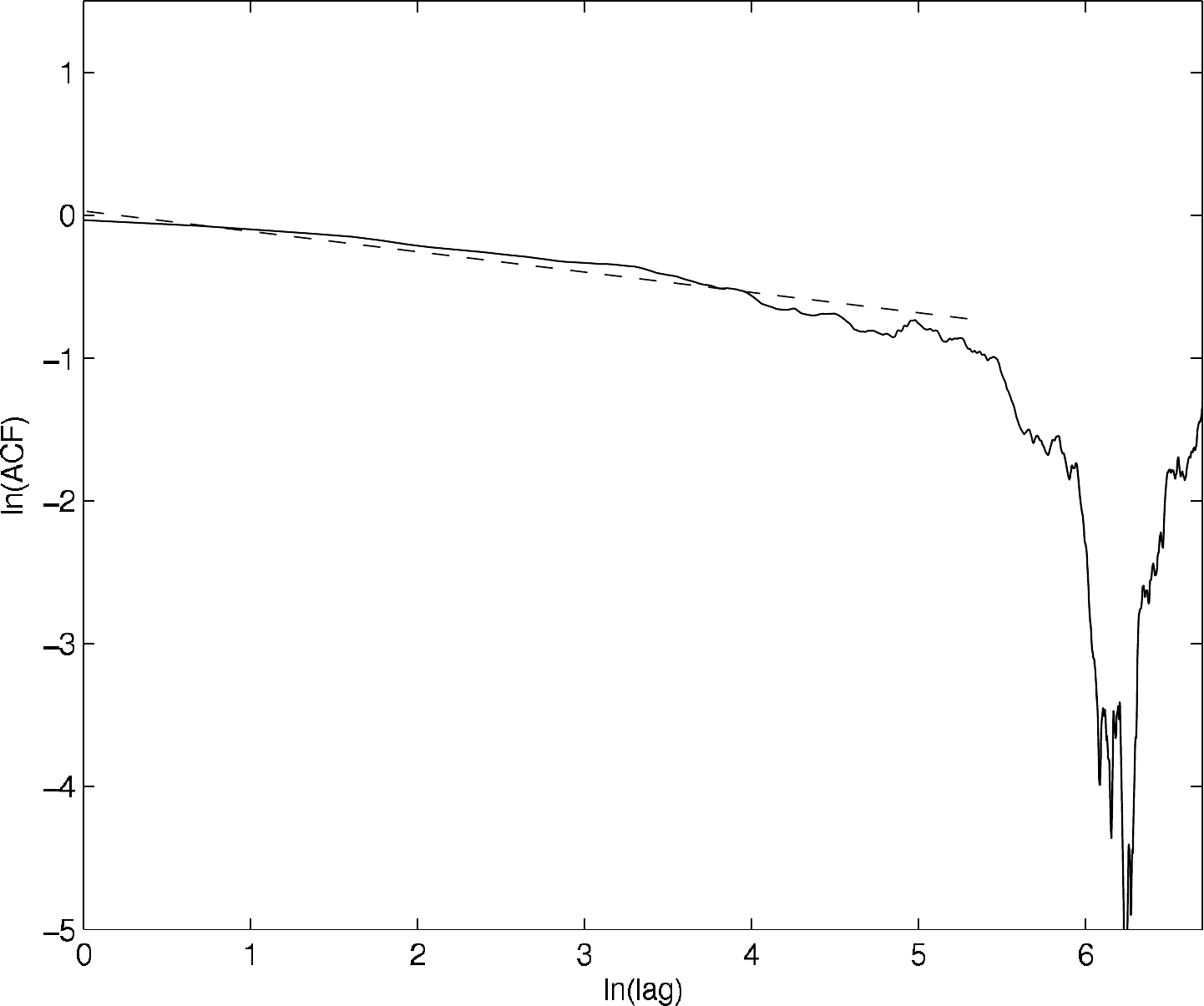}
\caption{ln-ln plot of ACF vs. time lag at pressure $1.4\times10^{-2}$ Torr. Up to 6 decorrelation times, it shows power law (dotted line), and after that it follows exponential decay.}
\label{fig:9}
\end{figure}

\begin{figure}[ht]
\centering
\includegraphics [width=8.5cm] {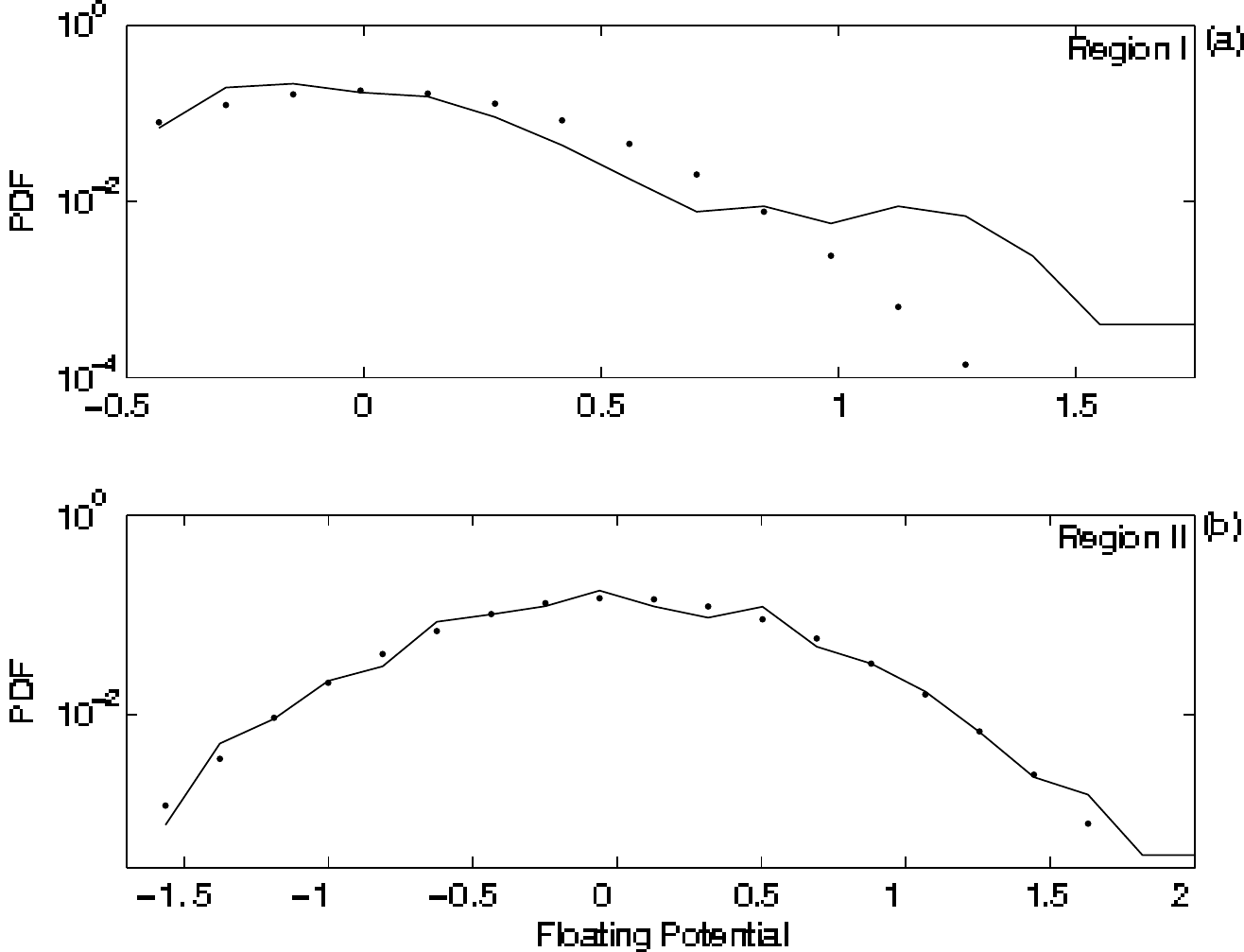}
\caption{The probability distribution function (PDF) of the fluctuation at pressure $1.2\times10^{-2}$ Torr, and the dotted line is the corresponding gaussian fit. }
\label{fig:10}
\end{figure}

Figure~\ref{fig:6} shows a typical plot of ln(R/S) vs. ln(time lag) of the fluctuations in region I, for a pressure of $1.4\times10^{-2}$. The Hurst exponent H is about $0.96\pm0.01$ (this indicates long-range time correlations)~\cite{PRL:BACarreras00}, and it is almost constant over the entire pressure range in region I.  On the other hand, for sinusoidal like oscillation in region II, the Hurst exponent is 1 for the lag length of one period of oscillations, and zero with more than one period lag~\cite{POP:carreras4} as shown in Figure~\ref{fig:7}. Also in the same region (II) multi slope ln(R/S) vs. ln(lag) plot as seen in Figure~\ref{fig:8} probably due to plasma instabilities of many frequencies~\cite{POP:carreras4}. The ACF exponent $\alpha$ has been calculated for the fluctuations in region I, from the ln(ACF) vs. ln(time lag) plot as shown in Figure~\ref{fig:9}. The ACF (Fig.~\ref{fig:9}) shows power law up to about 6 times the decorrelation time, and after that it follows exponential decay. Average value of $\alpha$ is about 0.30. Using the relation $H=(2-\alpha)/2$ ~\cite{PRE:GRangarajan}, H calculated from ACF is $\sim$ 0.85, which is close to the value of H, calculated using R/S technique. In region II, no power law decay of ACF has been observed. The PDF of the floating potential fluctuations in region I, seen in Fig.~\ref{fig:10}(a) clearly shows a nongaussian nature. Corresponding best gaussian fit is given by dotted curve in the same figure.

 We suspect that there might be a slightly bimodal distribution similar to Ref.~\cite{PhyslettA:Skokov}.  Figure~\ref{fig:10}(b) shows the gaussian nature of the fluctuations in region II.

Our results of Hurst exponent, $H>0.5$, ACF exponent, $\alpha\sim0.30$, nongaussian PDF, and power spectral index $\beta \sim 1.60$ in the pressure range $9\times10^{-3}-1.6\times10^{-2}$ Torr, are consistent with the systems exhibiting SOC like behavior.

Comparison of $\alpha$, H by ACF, H by R/S, and $\beta$ have been shown in  Table ~\ref{table1}, for pressures $0.9\times10^{-2}$, $1.2\times10^{-2}$, $1.5\times10^{-2}$ Torr.

\subsection{Conclusion}
\label{subsec:conclusion}
We have obtained SOC behavior over a finite range of neutral pressure of $ 9\times10^{-3}-1.6\times10^{-2}$ Torr for a fixed discharge voltage and this region belongs to the region less than the Paschen minimum~\cite{Physleta:nurujjaman}. Glow discharges are simple systems, but their physics can be quite complicated due to the presence of several phenomena like avalanche breakdown, ionization waves, low frequency ion-acoustic instability, double layer, chaos, etc. Most of them are highly nonlinear processes and hence one requires different techniques both statistical and spectral to investigate and understand their behavior. From our present analysis we observe that the plasma dynamics in the region I, is compatible with self organized criticality, while region II, is not. This could also imply that plasma transport in region I, is quite different from region II.

\section{Summary of the work}
\label{section:summary}

\subsection{Summary}
The thesis presents the experimental observations of the nonlinear phenomena$-$ chaos to
order transition, self organized criticality behavior and noise invoked processes like
the stochastic and coherence resonance in an argon dc glow discharge
 plasma [Fig~\ref{fig:setup} of Section~\ref{section:experiment}].
Depending upon the gas pressure (p), a discharge was struck at different discharge voltages (DV), which exhibits a Paschen curve like behavior [Fig.~\ref{fig:paschen} of Section~\ref{section:experiment}]. The region left to the Paschen minimum is narrower than the region to the right and the behavior of the plasma floating potential fluctuations were different on these two sides of the Paschen curve.

When the system was operated for the pressures greater than the Paschen minimum, an anode glow or anode spot was observed to form on the anode and with the increase in the DV its size decreased and finally diminished at a certain DV. The annular radii of the glow  estimated using CCD camera was in the range of $1.3-0.32$ mm.   This observation is very consistent with the theoretical estimations of thickness ($\delta$) using the relation $\delta\approx3.7\times10^{-6}\frac{kT}{\sigma_{i}P}$, where k, T, $\sigma_{i}$, and P are the Boltzmann constant, room temperature in Kelvin scale, ionization cross-section and pressure in mbar respectively. The estimated thickness of the anode glow using the above relation for $P=0.95$ mbar, $T=300$ K, is approximately 0.81 mm is  within the range of the thickness as obtained  from the experiment ($1.3-0.32$ mm).

An interesting feature associated with the anode glow was the different types of oscillations in the floating potential at different pressures, and these fluctuations have been analyzed for the three typical pressures in the region greater than the Paschen minimum. At about 0.89 mbar ($pd\approx 20.02$ mbar-mm), the discharge was initiated at $\approx 288$ V and simultaneously irregular relaxation type of oscillations were observed  and the regularity of the oscillation also increased with the DVs. Around 509 V,  these oscillations disappeared and this point (DV) is termed as the bifurcation point ($V_H$). Almost similar behaviour had been observed at 0.95 mbar ($pd=21.37$ mbar mm)  and at 1 mbar ($pd= 22.5$ mbar mm ). Initially, the power spectrum of the oscillations as of broad band nature indicating chaotic behavior of the system. The frequency of the oscillations were around the ion plasma frequency. An estimate of the frequency of these instabilities can be obtained from the ion transit time in the plasma $\tau(d)=\frac{d}{V_{th,i}}=d/\sqrt\frac{k_bT_i}{m}$, where d is the electrode distance. The estimated ion transit frequency ($\frac{1}{\tau}$) for our experimental system is $\approx19$ kHz which agrees well with the frequency of the relaxation oscillations of the floating potential.

 The presence of relaxation oscillations has been attributed to the formation of highly nonlinear structures like double layers. We therefore tried to estimate the correlation dimension ($D_{corr}$) and the +ve Lyapunov exponent ($\lambda_L$) of all the signals.  The estimated $D_{corr}$ for the data for the three pressures (0.89, 0.95 and 1.0 mbar) were greater than 3.8 to begin with and decreased with increase in DV. Since $D_{corr}$ is a measure of the complexity of the system, initially, the system was in a complex state as $D_{corr}$ for all the three pressures were high and decreased in complexity with increase in DV.

The presence of a +ve Lyapunov exponent ($\lambda_L$) is the most reliable signature of the chaotic dynamics.  The positive $\lambda_L$ has been identified  for DV  283, 284, and 290 V at 0.95 mbar and for 293, 296, 300 and 305 V at 1 mbar respectively.  At higher pressures we find that in general $\lambda_L$ is +ve and $D_{corr}  \geq 3$, suggesting a low dimensional chaos. Wavelet analysis also showed presence of chaos at the initial stage of the discharge.

Generally, the floating potential fluctuations were complex in nature at the initial stage of discharge and became regular with increase in DV and converted to relaxation oscillations and  vanished through homoclinic bifurcation at $V_H$. Applying noise beyond the $V_H$  noise invoked experiments, coherence and stochastic resonance had been performed. For the coherence resonance we had only applied noise  to the system to get back the autonomous dynamics, whereas, for the stochastic resonance, noise and periodic pulse had been applied simultaneously. In order to overcome the plasma related problem in analyzing the stochastic resonance, we had devised new analysis techniques absolute mean difference.

When the system was operated in the pressure region less than the Paschen minimum, for small range of p ($0.9-1.5\times10^{-2}$ Torr or $1.2-2\times10^{-2}$ mbar), it was observed that the behavior of the floating potential fluctuations was consistent with SOC. In order to establish the SOC behavior, we had checked the power law behavior of the power spectrum, the presence of the long-range correlation by estimating Hurst exponent (H) using R/S technique and the exponent ($\alpha$) of ACF decay, and the nongaussian probability distribution function. The results of Hurst exponent ($0.96\pm0.01$) greater than 0.5, ACF exponent, $\alpha\sim0.30$, nongaussian PDF, and power spectral index $\beta \approx 1.7\pm0.1$ in the pressure range $9\times10^{-3}-1.6\times10^{-2}$ Torr, are consistent with the systems exhibiting SOC like behavior.

\subsection{Scope of the future works}
Though we have investigated some of the nonlinear processes  during the the course of this thesis work, there are, many problems which we plan to take up as future plans. They are as follows.
\begin{itemize}
  \item Effect of the noise on the autonomous dynamics.
  \item Effect of suprathreshold signal and noise to the plasma.
  \item Investigation of the existence of canard orbits in plasma.
  \item Non-chaotic attractors may be investigated by applying two non-commensurate periodic signals.
      \item Chaos control and synchronization.
      \item Modeling of the experimental results.
\end{itemize}

Dusty plasma which can be produced very easily in our experimental system, is another area where, a lot of nonlinear phenomena  can be explored.

\section*{Aknowledgement}

        I wish to express my deep sense of respect and gratitude to my
thesis advisor, Prof. A.N. Sekar Iyengar who has been a constant source
of inspiration during the course of this thesis work. He has always been
patient towards my shortcomings and kept encouraging me to work in
a better way. He was never reluctant to venture out into new vistas
of research and would always motivate me to do so too. This freedom
in research has helped me express myself better and more importantly
always had a feeling of working independently. It has surely helped me to
become a better person.

        I would like to thank Prof Punit Parmananda for his many
suggestions and constant supports during the work `Noise invoked
dynamics in glow discharge plasma'.        I would like to thank Prof R.
Pal, Prof. N. R. Ray, Prof S.K. Saha, Prof M. S. Janaki, Prof. Nikhil
Chakrabarti, and Prof. S. Raychaudhuri for various suggestions, and
providing me various experimental diagnostics and materials during the
course of my thesis work.

        In order to carry out experiments in glow discharge plasma system,
one requires various technical knowhow. I would hence like to thank the
technical and scientific staff of the division, for providing me with various
assistances during the course of this thesis work. I thus acknowledge
the efforts of Amalendu Bal, S.S. Sil, Dipankar Das, Partha Sarathi
Bhattacharyya, Dipak Banik, Subhasis Basu, Monobir Chattapadhyay,
Santanu Chaudhuri, Ashok Ram and Abhijit Betal. A special thanks to
S.S. Sil and Dipaknar Das for their help in the mechanical part of the
experiments and Amalendu Bal for electronics circuitries. I also thank
our administrative staff Dulal Chaterjjee and previous administrative staff
Dipak Das for providing administrative assistances.

        I would also like to thank Prof. S. Banarjee of IIT Kharagpur and 
Prof S.K. Dana of Indian Institute of Chemical Biology, Kolkata, India, for
the useful discussions on nonlinear dynamics approach to plasma physics.

        A Ph.D. student's trauma and pressures are best understood by
fellow research scholars and seniors. Herein I would like to acknowledge
my seniors Ramesh Narayanan, Rajeev Kumar, Ramit Bhattacharyya,
Krishnendu Bhattacharyya, Debjyoti Basu and batch-mate Anirban Bose.
I would also like to thank my juniors Subir Biswas and Debobrata Banerjee
for giving me helping hands not only during my thesis work but also in
various other activities.

        Any developmental work in scientific institutions is not possible
without the assistance of the workshop. I therefore specially thank Jisnu
Da, Debnath Da, Rabin Da and Sudipto Da apart from the other workshop
staff for their active support regarding any developments required.

        During my carrier I am indebted to my school teachers, Naimuddin,
Nimai Mandal, Siddique Hossain, Nurul Hoda, Abdul Jabbar, Liakat Ali
and my uncle Najrul Islam, without their helping hands, I would not have
been able to come to this position.

        I also acknowledge my wife, Tazma, for her psychological support
during the PhD work. I also thank our kid Tanweer who has given us a
lot of cheers.

        Lastly, I pay my respects to the most important persons in my life,
`Father (late)' and `Mother' whose blessings and whole-hearted support
have helped me through all the ups and downs in my life.

%

\end{document}